\documentclass[
superscriptaddress,
nofootinbib,
amsmath,amssymb,
aps,
twocolumn,
preprintnumbers,
floatfix,
]{revtex4-2}

\usepackage{graphicx}
\usepackage{dcolumn}
\usepackage{bm}
\usepackage{xcolor}
\usepackage{wasysym}
\usepackage{physics}
\usepackage{subcaption}
\usepackage{soul}
\usepackage{cancel}
\usepackage{mathtools}
\usepackage{xcolor}
\usepackage{ragged2e}
\usepackage{tabularx}
\usepackage{array}
\usepackage[linkcolor=blue, colorlinks]{hyperref}

\definecolor{myGreen}{RGB}{17, 122, 17}
\definecolor{cmblue}{rgb}{0.12156862745098039, 0.4666666666666667, 0.7058823529411765}
\definecolor{mygrey}{gray}{0.35}
\definecolor{myblue}{rgb}{0.2,0.2,0.8}
\definecolor{mygreen}{rgb}{0.2,0.8,0.5}
\definecolor{myzard}{cmyk}{0,0,0.05,0}
\definecolor{mywhite}{rgb}{1,1,1}
\definecolor{myred}{rgb}{1,0.,0.3}
\definecolor{amethyst}{rgb}{0.6, 0.4, 0.8}

\begin{document}

\title{Real-Time Dynamics in a (2+1)-D Gauge Theory:\\ The Stringy Nature on a Superconducting Quantum Simulator}

\author{Jesús Cobos}
\email{jesus.cobos@ehu.eus}
\affiliation{EHU Quantum Center and Department of Physical Chemistry, University of the Basque Country UPV/EHU, P.O. Box 644, 48080 Bilbao, Spain}

\author{Joana Fraxanet}
 \affiliation{IBM Quantum, IBM Thomas J Watson Research Center, Yorktown Heights, NY 10598, USA}

\author{César Benito}
\affiliation{Instituto de Física Teórica, UAM-CSIC, Universidad Autónoma de Madrid, Cantoblanco, 28049 Madrid, Spain}
 
\author{Francesco di Marcantonio}
\affiliation{EHU Quantum Center and Department of Physical Chemistry, University of the Basque Country UPV/EHU, P.O. Box 644, 48080 Bilbao, Spain}
 
\author{Pedro Rivero}
 \affiliation{IBM Quantum, IBM Thomas J Watson Research Center, Yorktown Heights, NY 10598, USA}

\author{Korn\'el Kap\'as}
\affiliation{Strongly Correlated Systems Lend\"ulet Research Group, Wigner Research Centre for Physics, H-1525, Budapest, Hungary}

\author{Mikl\'os Antal Werner}
\affiliation{Strongly Correlated Systems Lend\"ulet Research Group, Wigner Research Centre for Physics, H-1525, Budapest, Hungary}

\author{\"Ors Legeza}
\affiliation{Strongly Correlated Systems Lend\"ulet Research Group, Wigner Research Centre for Physics, H-1525, Budapest, Hungary}
\affiliation{Institute for Advanced Study, Technical University of Munich,
Germany, Lichtenbergstrasse 2a, 85748 Garching, Germany}
\affiliation{Parmenides Stiftung, Hindenburgstr. 15, 82343, Pöcking Germany}

\author{Alejandro Bermudez}
\affiliation{Instituto de Física Teórica, UAM-CSIC, Universidad Autónoma de Madrid, Cantoblanco, 28049 Madrid, Spain}
 
\author{Enrique Rico}
\affiliation{EHU Quantum Center and Department of Physical Chemistry, University of the Basque Country UPV/EHU, P.O. Box 644, 48080 Bilbao, Spain}
\affiliation{DIPC - Donostia International Physics Center, Paseo Manuel de Lardizabal 4, 20018 San Sebastián, Spain}
\affiliation{IKERBASQUE, Basque Foundation for Science, Plaza Euskadi 5, 48009 Bilbao, Spain}
\affiliation{European Organization for Nuclear Research (CERN),  Theoretical Physics Department, CH-1211 Geneva, Switzerland}

\date{\today}

\preprint{CERN-TH-2025-111}

\begin{abstract}
Understanding the confinement mechanism in gauge theories and the universality of effective string-like descriptions of gauge flux tubes remains a fundamental challenge in modern physics. We probe string modes of motion with dynamical matter in a digital quantum simulation of a (2+1) dimensional gauge theory using a superconducting quantum processor with up to 144 qubits, stretching the hardware capabilities with quantum-circuit depths comprising up to 192 two-qubit layers. We realize the $Z_2$-Higgs model ($Z_2$HM) through an optimized embedding into a heavy-hex superconducting qubit architecture, directly mapping matter and gauge fields to vertex and link superconducting qubits, respectively. Using the structure of local gauge symmetries, we implement a comprehensive suite of error suppression, mitigation, and correction strategies to enable real-time observation and manipulation of electric strings connecting dynamical charges. Our results resolve a dynamical hierarchy of longitudinal oscillations and transverse bending at the end points of the string, which are precursors to hadronization and rotational spectra of mesons. We further explore multi-string processes, observing the fragmentation and recombination of strings. The experimental design supports 300,000 measurement shots per circuit, totaling 600,000 shots per time step, enabling high-fidelity statistics. We employ extensive tensor network simulations using the basis update and Galerkin method to predict large-scale real-time dynamics and validate our error-aware protocols. This work establishes a milestone for probing non-perturbative gauge dynamics via superconducting quantum simulation and elucidates the real-time behavior of confining strings.
\end{abstract}

\maketitle

\section{Introduction}

The emergence of a ``string-like'' nature in fundamental gauge theories has provided a paradigm for understanding universal non-perturbative aspects of the strong interaction. From the phenomenological linearity of rotational Regge trajectories in hadronic spectra~\cite{Regge1, ChewFrautschi} to the underlying framework for color confinement~\cite{Nambu2, Susskind1}, both anchored in the emergence of flux tubes between color charges, effective string-like models can faithfully capture the low-energy properties of the confinement regime. These models extend beyond the leading-order Nambu-Goto string~\cite{LuscherWeisz, Luscher1, PolchinskiStrominger}  to include universal corrections derived from the finite width, rigidity, and coupling of the flux tube to massive modes. Such a string-like picture has received strong support from lattice gauge theories (LGTs), which have provided ab initio evidence for the formation of the confining color flux tubes connecting static quarks. These studies have reproduced the linear confining potential \cite{Bali1, JugeKutiMorningstar, gliozzi2010width} and other static properties, such as the tension of the flux tube, its width, and possible excited states. 

The situation is very different for real-time dynamics, where the sign problem in Monte Carlo simulations and the entanglement barrier in tensor network methods have hindered a similar progress. Quantitative ab initio studies of real-time string formation, longitudinal and transverse string motion intertwined with dynamical charges, as well as string fragmentation and recombination reaching long thermalization timescales, are still missing for (3+1)-D quantum chromodynamics (QCD).

\begin{figure*}
\includegraphics[width=1.0\textwidth]{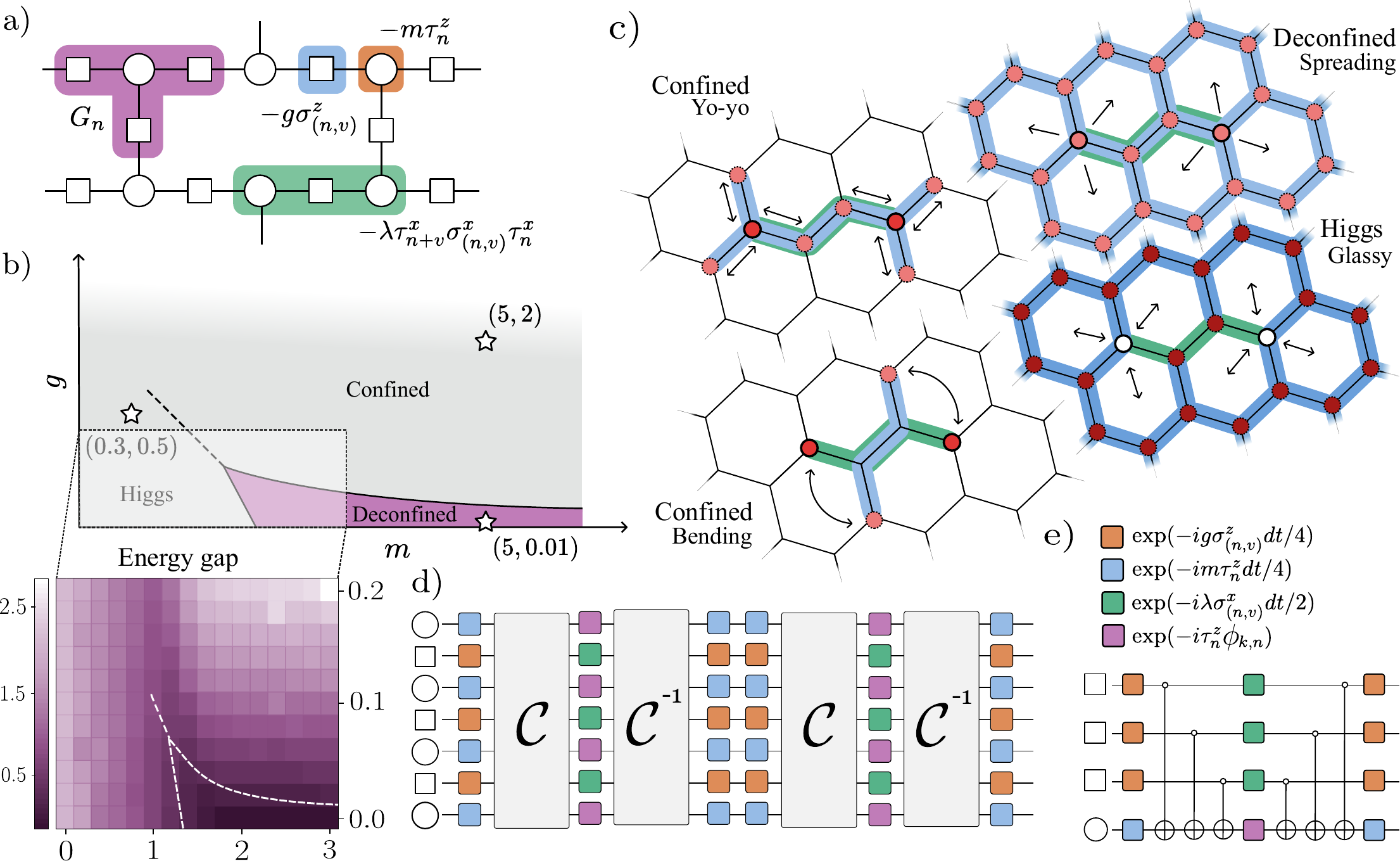}
\caption{\justifying Outline of the $Z_2$HM. (a) Shows a sketch of the support of the different terms of Hamiltonian \eqref{eq:hamiltonian} and the gauge transformation operators $G_n$. In (b), we sketch the phase diagram of the model and present data for the energy gap from large-scale density matrix renormalization group (DMRG) computations. We use stars to highlight the value of the microscopic parameters $(m, g, \lambda)$ used for the real-time quantum and MPS simulations sketched in (c). We consider three distinct sets of values for these $\{(5, 2, 1), (5, 0.01, 1), (0.3, 0.5, 1)\}$, corresponding to each of the static phases of the model, and find three dynamical regimes. The structure of the Trotter circuits implementing the real-time evolution is displayed in (d). These quantum circuits are built by repeated, ordered composition of the Pauli gadget depicted in (e). $\mathcal{C}$ is a dense block of CNOT gates with depth 3, which are grouped after commutation. The two-qubit depth of these circuits is $D=6N_eL$ for $N_e$ edges on the simulated lattice and $L$ Trotter depth.}   
\label{fig:fig0}
\end{figure*}

Although QCD has proven to be the fundamental theory of strong interactions, simplified models have long been a powerful tool for isolating and understanding common driving mechanisms. In gauge theory, models such as the (2+1)-D $Z_2$HM can capture key confinement features and string dynamics within a more tractable setting~\cite{Kogut1, PhysRevD.19.3682}. More importantly, they unveil a unique opportunity, bringing the direct experimental observation of non-perturbative string-like dynamics within reach of current quantum platforms. This shifts the focus from asymptotic in-out probes in collider experiments to the real-time microscopy of flux tubes and their non-equilibrium evolution in table-top quantum devices~\cite{banuls2020simulating, bauer2023quantum, di2024quantum}. Instead of using a semiclassical phenomenological description of hadronization \cite{andersson1998lund}, which aims to predict the asymptotic hadron yields observed in colliders by treating string dynamics classically and reserving quantum effects for pair production, direct observation of string dynamics opens the possibility of quantitative tests with pure quantum treatment.

This article presents a significant step in this direction, reporting on the experimental observation of the real-time dynamics of electric strings in a (2+1)-D $Z_2$HM. Leveraging recent advancements in superconducting circuits and integrating a comprehensive toolbox of error suppression, mitigation, and correction techniques,  we have realized the dynamics of this LGT and tested string-like physics in regimes that have required stretching the capabilities of superconducting quantum hardware to the limit. This has enabled the application of tailored quench protocols to excite electric strings, the subsequent measurement of their propagation with real-time resolution, and the exploration of new multi-string scenarios. The content of this work aligns with recent theoretical proposals on quantum simulation of LGTs~\cite{marcos2014two,gustafson2021toward,lumia2022two,homeier2023realistic,irmejs2023quantum} that address the experimental implementation of such models. In contrast, actual implementations in (2+1)-D \cite{cochran2024visualizing,gonzalez2025observation}, where richer topological and dynamical effects appear, remain comparatively scarce.

We take a step forward and use IBM superconducting chips to track how the string stretches in longitudinal ``yo-yo" modes and how it also exhibits transverse ``wiggling" localized near its endpoints. We identify key differences between the quantum and semiclassical realms in these precursors to rotational dynamics and Regge trajectories of meson-like composites. Going beyond a single string, we demonstrate how multi-string configurations can fragment and reorganize in a heavy-massive regime, distinct from conventional string breaking via particle-antiparticle creation. Our observations, validated by tensor network simulations, provide a direct bridge between the theoretical constructs of effective string-like models and tangible, dynamic observables, opening a new frontier for probing the non-perturbative physics of gauge theories. Moreover, the methods developed and the lessons learned are broadly applicable to quantum simulations beyond the specific model studied in this work.

\section{The $Z_2$-Higgs model \& static properties}

Gauge-Higgs models are central in understanding confinement and symmetry breaking in LGTs~\cite{Kogut1,PhysRevD.19.3682}. In the $Z_2$ case \cite{bonati2022multicritical,borla2022quantum}, paralleling the situation with other gauge groups, the Higgs and confined phases are not separated by a sharp phase transition but are instead smoothly connected. In the square lattice, which has also proven to be foundational in condensed matter~\cite{PhysRevB.82.085114}, the deconfined phase underlies the topological order in some types of quantum spin liquids~\cite{KITAEV20032}, while charge or flux condensation leads to the confined and Higgs phases, respectively.

We adapt the $Z_2$HM to a particular hardware, IBM superconducting chips with heavy-hexagonal connectivity,  used to minimize frequency collisions for high-fidelity  gates~\cite{PhysRevLett.107.080502,PhysRevX.10.011022}. The model hereby realized is a LGT with Pauli matter and gauge fields with a Hamiltonian,
\begin{equation}
H = -m \sum_{\boldsymbol{n}} \tau_{\boldsymbol{n}}^z - g \sum_{(\boldsymbol{n},  \boldsymbol{v})} \sigma_{(\boldsymbol{n},  \boldsymbol{v})}^{z} - \lambda \sum_{\boldsymbol{n}, \boldsymbol{v}} \tau_{\boldsymbol{n} + \boldsymbol{v}}^x \sigma_{(\boldsymbol{n}, \boldsymbol{v})}^x \tau_{\boldsymbol{n}}^x.
\label{eq:hamiltonian}
\end{equation}
Here, $\tau$, $\sigma$ are Pauli operators defined in a constrained tensor-product Hilbert space, $\boldsymbol{n}$ denotes the sites of the hexagonal lattice, while $\boldsymbol{v}$ stands for the unit lattice vectors; in the basis where $\tau^z$, $\sigma^z$ are diagonal, we define their eigenvectors by $\tau^z |0\rangle = |0\rangle$, $\sigma^z |0\rangle = |0\rangle$, and $\tau^z |1\rangle = -|1\rangle$, $\sigma^z |1\rangle = -|1\rangle$. As usual, matter fields live on the sites and gauge fields on the links, as shown in Fig.~\ref{fig:fig0}(a). The first two terms $H_M = m\sum_{\boldsymbol{n}} \tau_{\boldsymbol{n}}^z$ and $H_E = g \sum_{\boldsymbol{n}, \boldsymbol{v}} \sigma_{(\boldsymbol{n}, \boldsymbol{v})}^z$, encode the local energies of matter and electric fields, while $H_I = \lambda \sum_{\boldsymbol{n}, \boldsymbol{v}} \tau_{\boldsymbol{n} + \boldsymbol{v}}^x \sigma_{(\boldsymbol{n},\boldsymbol{ v})}^x \tau_{\boldsymbol{n}}^x$ defines their gauge-invariant coupling.

Note that this Hamiltonian differs from the traditional Kogut-Susskind Hamiltonian for LGTs~\cite{PhysRevD.11.395}, as magnetic plaquette terms inducing direct fluctuations of electric field configurations are absent. In doing this, we avoid a considerable circuit-depth overhead in a Trotter expansion since the plaquette term would require a six-body interaction in the heavy-hex lattice. We emphasize, however, that this does not preclude resolving the phenomenology of a deconfined phase, as dynamical matter can tunnel along closed loops and lead to effective plaquette fluctuations, as discussed in more depth in the following. The $Z_2$ gauge symmetry is generated by the operators, 
\begin{equation}
G_{\boldsymbol{n}} = \tau_{\boldsymbol{n}}^z \prod_{\boldsymbol{v }\in \ell_{\boldsymbol{n}}} \sigma_{(\boldsymbol{n}, \boldsymbol{v})}^z, \hspace{0.8cm} \left[G_{\boldsymbol{n}}, H\right] = 0 \hspace{0.5cm} \forall \boldsymbol{n},
\label{eq:gauge}
\end{equation}
where $\ell_{\boldsymbol{n}}$ denotes the directions of the links connected to site $\boldsymbol{n}$ in this trivalent lattice (see Fig.~\ref{fig:fig0}(a)). Since gauge symmetries commute with the Hamiltonian, they are constants of motion,  dividing the complete Hilbert space into sectors with different eigenvalues $G_{\boldsymbol{n}} \ket{\psi} = \pm \ket{\psi}.$ These are related to the absence ($+$) or presence ($-$) of a static background charge at the site $\boldsymbol{n}$. We focus on physical states stabilized by the generators, 
\begin{equation}
G_{\boldsymbol{n}} \ket{\psi} =  \ket{\psi},
\label{eq:physical_states}
\end{equation}
which can be understood as a discrete Gauss' law.

This model has three distinct regimes sketched in Fig.~\ref{fig:fig0}(b). The Higgs regime appears for small values of $m$ and $g$, while the confined regime emerges when both $m$ and $g$ are sufficiently large. Despite the absence of a plaquette term, a deconfined phase appears at large $m$ and very small $g$. In the Higgs regime, the ground state of the model is a highly entangled non-local superposition of all the physical states in the eigenbasis of $H_M$ and $H_E$, which correspond to the classical configurations of the matter and gauge fields. In limits $m\to 0$ or $g \to 0$, the model can be diagonalized in terms of mutually commuting stabilizer operators $\tau_{\boldsymbol{n} + \boldsymbol{v}}^x \sigma_{(\boldsymbol{n}, \boldsymbol{v})}^x \tau_{\boldsymbol{n}}^x$ and $\tau_{\boldsymbol{n}}^z \prod_{\boldsymbol{v} \in \ell_{\boldsymbol{n}}} \sigma_{(\boldsymbol{n}, \boldsymbol{v})}^z$. The ground state is non-degenerate, and there is a finite energy gap between the ground and first excited states. 

In the confined regime, the eigenstates are close in energy and fidelity to those of $H_M$, $H_E$, the ground state approaches $\ket{000\dots0}$ as $m$ or $g$ increases, and gauge-invariant excitations correspond to localized matter charges connected by electric field lines. The large value of $m$ leads to a global $U(1)$ symmetry related to the conservation of the total number of charges. Gauge invariance forces the pairs of charges to be connected by an electric field string, which has a large energetic cost proportional to $g$ to stretch or compress, and yields an effective potential growing linearly with their relative distance. In this phase, the mean local matter magnetization takes the value $\langle \tau^z \rangle \simeq 1$, as shown in Fig.~\ref{fig:phasediag_numerics} of the extended data.

The confined and Higgs regimes are adiabatically connected, since we find no gap closing as the microscopic parameters are varied; see Fig.~\ref{fig:fig0}(b). This characteristic is maintained in the thermodynamic limit for small values of $m$, and it follows from the fact that the confined ground state is contained in the superposition of the Higgs ground state; thus, no energy crossings or gap closure occur along the adiabatic path. For $m = 0$, the Hamiltonian becomes the sum of commuting terms,
\begin{equation}
H = -\sum_{\boldsymbol{n}, \boldsymbol{v}} \left( g\,\sigma_{(\boldsymbol{n}, \boldsymbol{v})}^z + \lambda \, \tau_{\boldsymbol{n}+\boldsymbol{v}}^x \sigma_{(\boldsymbol{n}, \boldsymbol{v})}^x \tau_{\boldsymbol{n}}^x \right),
\label{eq:hamiltonian_m0}
\end{equation}
which can be simultaneously diagonalized in the absence of the gauge constraint \eqref{eq:physical_states}. We can find the energy gap when reinstating the constraint, 
\begin{equation}
\Delta E = 2\sqrt{g^2 + \lambda^2},
\end{equation}
as described in Sect.~\ref{supp:m_0_analytics} of the supplemental material. Notice that $\Delta E$ is a strictly increasing function of $g$, while the ground state smoothly transforms to the $|000\dots000 \rangle$ state for $g \gg 1, m = 0$, which proves that the Higgs and confined regimes are adiabatically connected at least through the $m = 0$ axis of the phase diagram.

The deconfined phase appears for large $m$ and small $g$, and it is characterized by a vanishingly small energy gap, as shown in the large-scale DMRG computations of Fig.~\ref{fig:fig0}(b). Remarkably, this phase is found without an explicit plaquette term and is driven instead by the interplay between gauge fields and dynamical matter. The consecutive off-resonant tunneling through virtual charges effectively yields a plaquette operator at sixth order. We can thus estimate the extension of a deconfined region dominated by electric field fluctuations for $g \lesssim J_{\hexagon}^{\mathrm{eff}} = \gamma \lambda^6 m^{-5}$, where $J_{\hexagon}^{\mathrm{eff}}$ is the effective plaquette coupling. We numerically estimate $\gamma\approx0.25$ from the gap of a single hexagon for $m \gg \lambda$ and $g=0$. Hence, the induced magnetic flux terms dominate in this phase, and the ground state tends to a superposition of all the possible configurations of closed electric field strings with no matter excitations as $g/J_{\hexagon}^{\mathrm{eff}}\to 0$.

\begin{figure*}[ht]
\includegraphics[width=0.99\textwidth]{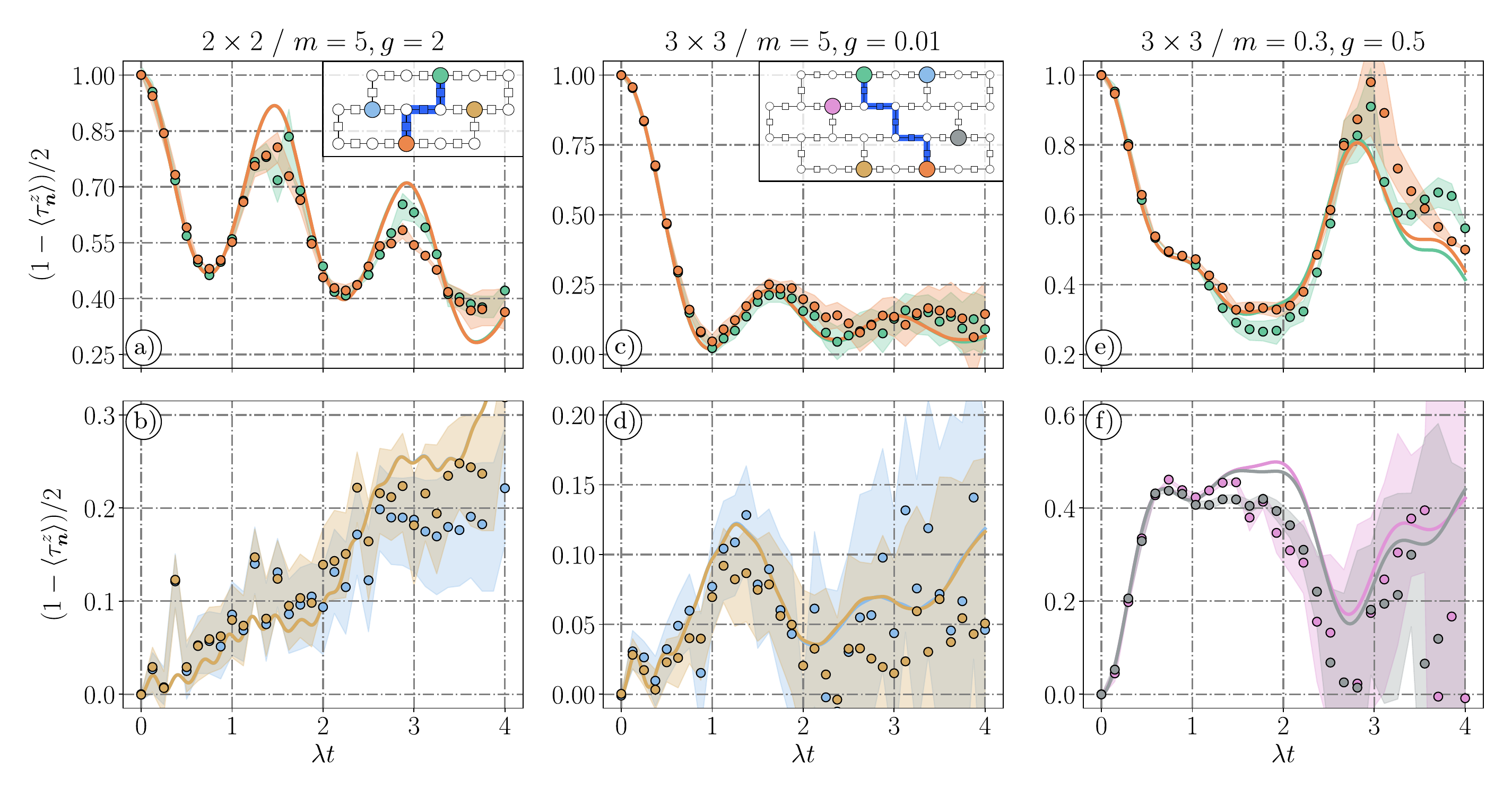}
\caption{\justifying Single string dynamics at different points of the phase diagram. (a) Features the occupation at the initial endpoints of the string in the confined phase in a $2\times2$ lattice (35 qubits) and Trotter $dt=0.15$ (2280 two-qubit gates). The yo-yo and bending modes are distinguished as short-period oscillations and a steady decrease in the mean occupation, respectively. The occupation in the endpoints of the rotated strings in this regime is shown in (b). Here, only bending and vacuum fluctuations are present. The colors of the curves indicate the site where the local occupation operators are measured. (c-d) show the dynamics of the occupation in the deconfined phase in a $3 \times 3$ lattice (68 qubits) with $dt=0.125$ (4286 two-qubit gates), where matter spreads all over the lattice and the system reaches a quasi-stationary state. (e-f) show the dynamics in the Higgs phase for the $3 \times 3$ lattice and $dt=0.15$. In this regime, the local occupations present a long-lived, damped oscillating behavior. Shaded regions indicate $70\%$ bootstrapping confidence.}
\label{fig:single_string_dynamics}
\end{figure*}

\section{Quench dynamics} \label{sec:results}

Let us now move to real-time dynamics, focusing on quantum quenches after a sudden change in the microscopic Hamiltonian couplings. In this section, we experimentally explore the quench dynamics for the $Z_2$HM, and provide quantitative tests with classical MPS-based simulations that provide a full quantum-mechanical treatment of string dynamics. In every experimental run, we consider the $\lambda = 0$ eigenstates of the Hamiltonian \eqref{eq:hamiltonian} as the initial state. These are configurations with well-defined charge positions, which appear in pairs connected with a string of electric field. We simulate this quench dynamics on \texttt{ibm\_kingston} and \texttt{ibm\_marrakesh}, starting from two distinct initial configurations, containing two (1-string) or four (3-string) charges, and evolving with the $\lambda = 1$ Hamiltonian at multiple values of $m, g$, as shown in Figures~\ref{fig:single_string_dynamics} and \ref{fig:double_string_dynamics}. We observe three dynamical regimes in correspondence to the static ground states discussed previously and predicted by the DMRG computations of Sect.~\ref{subsec:methods_dmrg}.

To obtain the experimental results, we have dealt with two sources of error: Trotter error and hardware noise, which we quantify in Fig.~\ref{fig:error_mitigation_data}. The first is the algorithmic error associated with approximating the dynamics of the model with a digital quantum circuit of finite depth. Due to the limited depth that current machines can reliably achieve, we cannot formally prove using existing Trotter error bounds that our circuits have sufficient depth to correctly reproduce the desired dynamical time intervals. However, we have found that these bounds are not tight for the problem at hand, since we do verify the consistency of our results using exclusively data from the quantum device by studying the same timescales with different Trotter time steps $dt$,  checking that there is no appreciable difference (see Fig.~\ref{fig:error_mitigation_data}). We further describe the Trotter circuits and algorithmic error in the supplemental material (Sect.~\ref{subsec:methods_trotter}). 

Hardware noise, on the other hand,  is the most restrictive error source in practice. To deal with it, we use a family of noise suppression, mitigation, and correction techniques, integrating novel strategies that we call gauge dynamical decoupling and Gauss sector correction, as well as previously known techniques such as Pauli twirling \cite{geller2013efficient} and operator decoherence renormalization \cite{farrell2024scalable}. In Methods Sect.~\ref{subsec:methods_em}, we describe these crucial techniques and address their performance and synergies. In Fig.~\ref{fig:error_mitigation_data}, we provide quantitative data supporting this discussion. We compare quantum quench simulations with MPS-based numerical methods, using the state-of-the-art ``basis update and Galerkin" numerical integrator~\cite{Ceruti-2023}, outlined in Sect.~\ref{supp:methods_mps_dyn} of the supplemental material. In Sect.~\ref{subsec:methods_execution}, we provide details about the execution of the simulations on IBM's quantum hardware, such as the number of shots used and the execution time. To quantify the errors in the quantum simulations, we apply bootstrapping, see Sect.~\ref{boots}, a non-parametric resampling technique that allows us to estimate the statistical uncertainty of observables derived from quantum measurement data.

\subsection{Single string dynamics}

To describe the 1-string quench dynamics of the $Z_2$HM, we measure the expectation value of local occupation operators at different matter sites. Local expectation values are less affected by the propagation of errors in current NISQ devices, yet they enable us to extract valuable physical information from IBM's quantum simulators. We performed three simulations with different values of the microscopic parameters, one in each of the parts of the phase diagram: confined, deconfined, and Higgs. The results are presented in Fig.~\ref{fig:single_string_dynamics}.

\subsubsection{String dynamics in the confined phase}

In the confined phase of the $Z_2$HM, matter particles and strings remain considerably localized after quenching, and the dynamics can be described using these as fundamental objects. In Figs.~\ref{fig:single_string_dynamics}(a)-(b), we experimentally observe some characteristic features of confinement in the presence of dynamical matter. The effective linear potential between matter particles yields dynamics which can be intuitively understood as the quantum analog of two masses connected by an elastic string, in correspondence with phenomenological descriptions. The charges in the initial state of one string have an undefined momentum, and all possible modes of motion of the string are initially excited in the dynamics. In the simulations, we observe two distinct modes: \textit{yo-yo} and \textit{bending}, corresponding to first- and second-order transitions, respectively. The yo-yo is a fast longitudinal oscillating mode that affects the position of particle pairs relative to their center of mass, causing matter particles to tunnel back and forth periodically with the consequent stretching and compressing of the string sketched in Fig.~\ref{fig:fig0}(c). This corresponds to the oscillations observed in the occupation of the initial string endpoints for the short timescales of the simulation shown in Fig.~\ref{fig:single_string_dynamics}(a). The population lost at these sites is transferred to the neighboring ones, as shown in Fig.~\ref{fig:supp_1string_otherp}(a)-(b) of the extended data. 

The bending mode of motion refers instead to the displacement of matter particles in the initial pair to positions where the final length of the electric field string remains unchanged. It is manifest in the slower decrease in occupation that accompanies yo-yo oscillations in Fig.~\ref{fig:single_string_dynamics}(a), and in the steady increase in occupation of matter sites in the rotated string positions in the lower panel of Fig.~\ref{fig:single_string_dynamics}(b). The small oscillations predicted by the MPS simulation of Fig.~\ref{fig:single_string_dynamics}(b) correspond to vacuum matter fluctuations with small amplitude due to the large value of $m$. Although fragmentation in the Lund string model is primarily driven by the classical yo-yo motion together with quantum pair creation,  the transverse bending near the string endpoints also plays a role, as it influences the angular distribution of emitted hadrons.  In our full quantum simulations, the large values of $m$ and $g$ forbid this fragmentation, and we can see how the bending eventually translates, for longer times, into a rotation of the string. However, the absence of plaquette terms in Eq.~\eqref{eq:hamiltonian} endows the string with further stiffness, making this mode slower and, for larger string sizes, only allowing for a partial transverse reorganization. For the depths allowed by the hardware, we can resolve the initial instants of this rotation, in which the string endpoints move to the closest lattice sites respecting the length of the string, as sketched in Fig.~\ref{fig:fig0}(c). On the smaller lattice that we have used (34 qubits), this is equivalent to one-quarter of the complete rotation of the string. Our choice of parameters $m = 5\lambda$, $g = 2\lambda$ induces a separation of the characteristic timescales for each of these modes of motion, allowing us to distinguish a faster yo-yo and a slower bending, which shows a remarkable agreement with the MPS-based simulations thanks to our integrated error-aware protocols.

To examine this dynamics more closely, we note that the initial 1-string state sketched in the inset of Fig.~\ref{fig:single_string_dynamics}(a) is close in energy and fidelity to a reduced number of eigenstates of~\eqref{eq:hamiltonian} with similar particle numbers and string lengths. Hence, these are coupled by the dynamics at low interaction orders. Since $m$ is large, the Hamiltonian has an approximate $U(1)$ symmetry, and transitions involving tunneling of matter particles are favored over particle creation from the vacuum. The value of $g$, which is large compared to $\lambda$, favors transitions in which the change in the length of the electric field string connecting the two matter particles is minimal. In this setting, the frequency of the yo-yo oscillations is $\omega_{\mathrm{y}} = 2g$. For $g = 2$, as in Fig.~\ref{fig:single_string_dynamics}, this leads to an oscillation period of $T_\mathrm{y} = 2\pi/\omega_{\mathrm{y}} \simeq 1.57 \lambda^{-1}$, which is in good agreement with both the experiment and the MPS simulations. 

We note that the discreteness of space in lattice models forces the bending to be a second-order effect. The initial and final states of the bending transitions have the same energy, but this is not the case for the intermediate states, which have smaller or larger string lengths. The characteristic frequency of the bending mode is estimated from perturbation theory to be $\omega_{\mathrm{b}} = \lambda^2/g - \lambda^2/(2m+g)$, which is considerably lower than that of the yo-yo in the confined regime. The two contributions arise from two different second-order transition paths in which two interaction operators act: one on the last bond of the original string and one on the last bond of the new configuration. The two paths differ in the order of these operations and have intermediate states with (unperturbed) energies that differ by $-2g$ and $2g + 4m$ from the initial state, which also leads to the relative sign in their contribution to the bending mode frequency. The noise in the device prevents us from resolving a complete bending oscillation with period $T_{\mathrm{b}} = 2\pi/\omega_b \simeq 15.08\lambda^{-1}$. In Fig.~\ref{fig:supp_1string_tn} of the extended data, we present MPS-based longer-time simulations that enable the resolution of more than half a bending oscillation.

\subsubsection{Matter spreads in the deconfined phase}

The deconfined phase appears for large $m$ and $g < J_{\hexagon}^{\mathrm{pert}}$, where the number of charges is approximately conserved and there are large electric-field fluctuations. Contrary to what happens in the confined phase, these charges can now spread without the electric energy cost through the lattice because the energy to change the length of a string, which is controlled by $g$, is no longer dominant. In the initial location of the matter particles whose occupation is shown in Fig.~\ref{fig:single_string_dynamics}(c), we observe a sudden population loss caused by the delocalization of the charges into initially empty sites, such as those shown in Fig.~\ref{fig:single_string_dynamics}(d). After this quick process, we observe that the occupation in all matter sites acquires a damped oscillating behavior around $\langle n \rangle \simeq 0.1$. This value is close to $2/N_n \simeq 0.07$ expected if the particle number is strictly conserved. These dynamics are characteristic of hard quenches in which the initial state has low fidelity with the eigenstate of the evolution Hamiltonian with the closest energy, and have been studied in one-dimensional spin systems~\cite{PhysRevB.102.054444}.

The spectrum of Hamiltonian \eqref{eq:hamiltonian} in the deconfined phase has the following structure: States with the same number of matter particles and distinct configurations of the gauge field share similar energies, while there is a large gap between state manifolds with different particle numbers. This implies that the quench dynamics is governed by transitions between states on the manifold with the closest energy to that of the initial state, in our simulations, the 2-particle sector. The density of states of this manifold is large, which implies that a long-lived superposition of these states is quickly reached after some oscillatory behavior characteristic of coherent evolution. This situation is not equivalent to reaching an equilibrium state, although the phenomenology is similar for the short times considered in this simulation. The finite dimensionality of the 2-particle sector implies that a revival of the initial configuration is expected after some long time $t_\mathrm{R}$, which increases rapidly with the size of the system. The amplitude of these revivals decreases with time and system size, and we expect thermalization to occur in this regime, following from non-integrability. However, the characteristic time scale of this phenomenon, in this particular model with local interactions, surpasses the capabilities of current quantum and classical simulations for these system sizes.

\subsubsection{Damped glassy oscillations in the Higgs phase}

The quench dynamics in the Higgs regime arise from an interplay between glassy oscillations, which dominate for small $m$, and matter spreading as in the deconfined phase that becomes manifest as $m$ grows, as shown in Fig.~\ref{fig:single_string_dynamics}(c)-(d). Here, the term glassy refers to the fact that the Hamiltonian for $m=0$, $g=0$ has an exponentially degenerate ground state subspace from which the Gauss law selects a disordered state in which all the local expectation values are zero, as sketched in Fig.~\ref{fig:fig0}. In the completely glassy regime $m = 0$, the dynamics can be solved analytically due to the commutativity of the terms in Hamiltonian \eqref{eq:hamiltonian_m0}. For the initial state considered, the expectation value of the local matter and gauge energies is, respectively $\langle \tau_{\boldsymbol{n}}^z(t) \rangle = a(t)^{d_{\boldsymbol{n}}} \langle \tau^z_{\boldsymbol{n}}(0) \rangle$, $\langle \sigma_{({\boldsymbol{n}}, {\boldsymbol{v}})}^z (t) \rangle = a(t)\langle \sigma_{({\boldsymbol{n}}, {\boldsymbol{v}})}^z (0) \rangle$, with
\begin{equation}
a(t) = 1 - \frac{2\lambda^2}{\lambda^2 + g^2} \, \sin^2 \left( t\sqrt{\lambda^2 + g^2} \right),
\end{equation}
and $d_{{\boldsymbol{n}}}\in\{2,3\}$ the degree of node $n$ of the lattice. Since the dynamics couples two states with opposite values of gauge degrees of freedom, as $g$ grows, the transition from the initial state to the second configuration is energetically inhibited, as the initial state approaches an eigenstate of the Hamiltonian, and the quench dynamics disappears. In the $m=0$ limit, the system does not thermalize. For finite values of $m>0$, the creation of charges from the vacuum is also suppressed, and charge spreading is favored as in the deconfined phase. This causes the glassy oscillations to damp, as in Figs.~\ref{fig:single_string_dynamics}(e)-(f), until the system eventually thermalizes after some revivals. The damping rate of these oscillations is proportional to $m/g$. 

\subsection{3-string dynamics}

\begin{figure}
\includegraphics[width=0.49\textwidth]{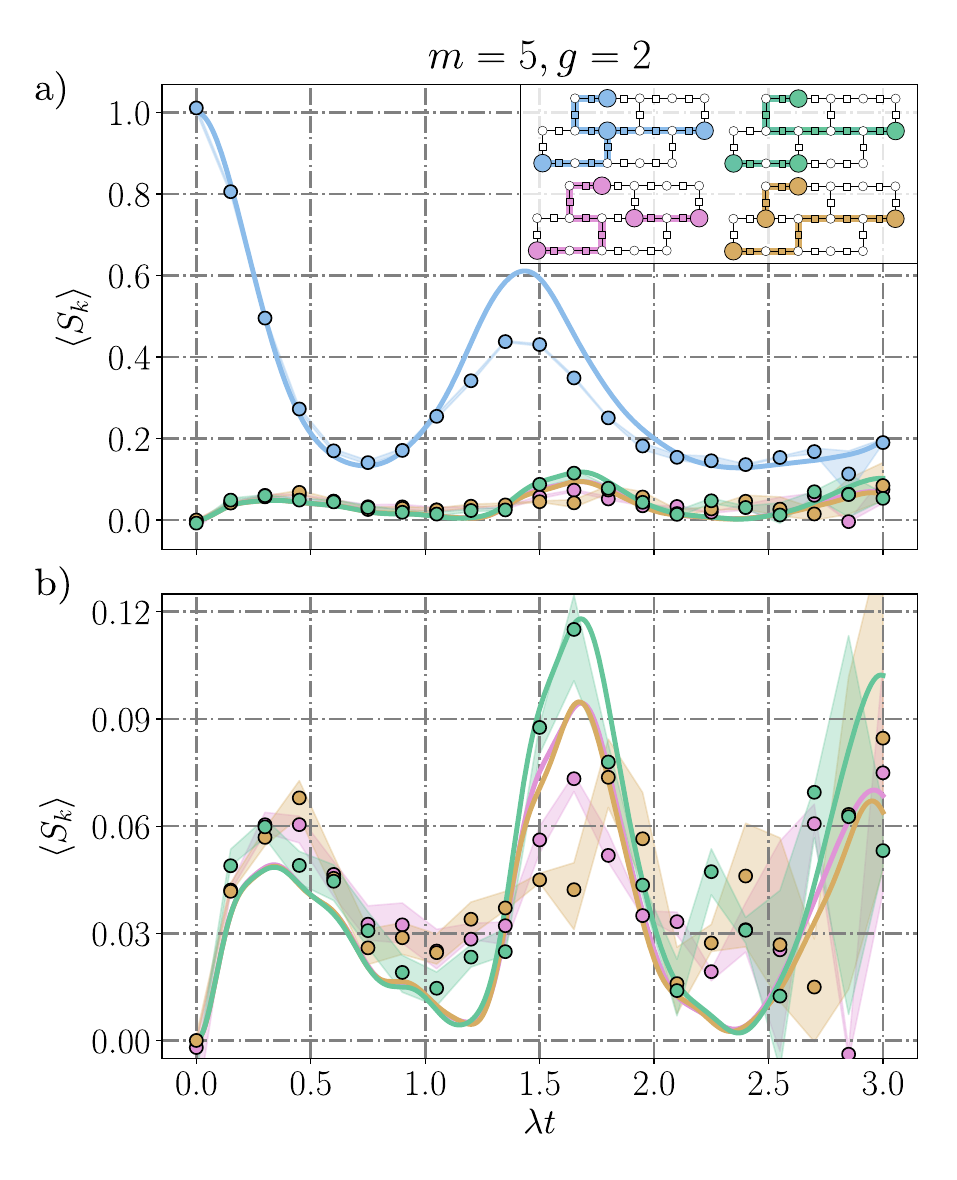}
\caption{\justifying Dynamics of string-like correlators $\langle S_k \rangle$ for the initial 3-string and broken string configurations in a $2\times2$ lattice in the confined phase $m = 5$, $g=2$. For each configuration $k$, $S_k$ is defined as the product of occupation operators in the matter sites indicated in the inset of (a). These operators quantify the population of each configuration. In (b), we highlight the population of the broken string configurations. We set $dt=0.125$ for the Trotter circuits.}
\label{fig:double_string_dynamics}
\end{figure}

\begin{figure*}[!ht]
\includegraphics[width=\textwidth]{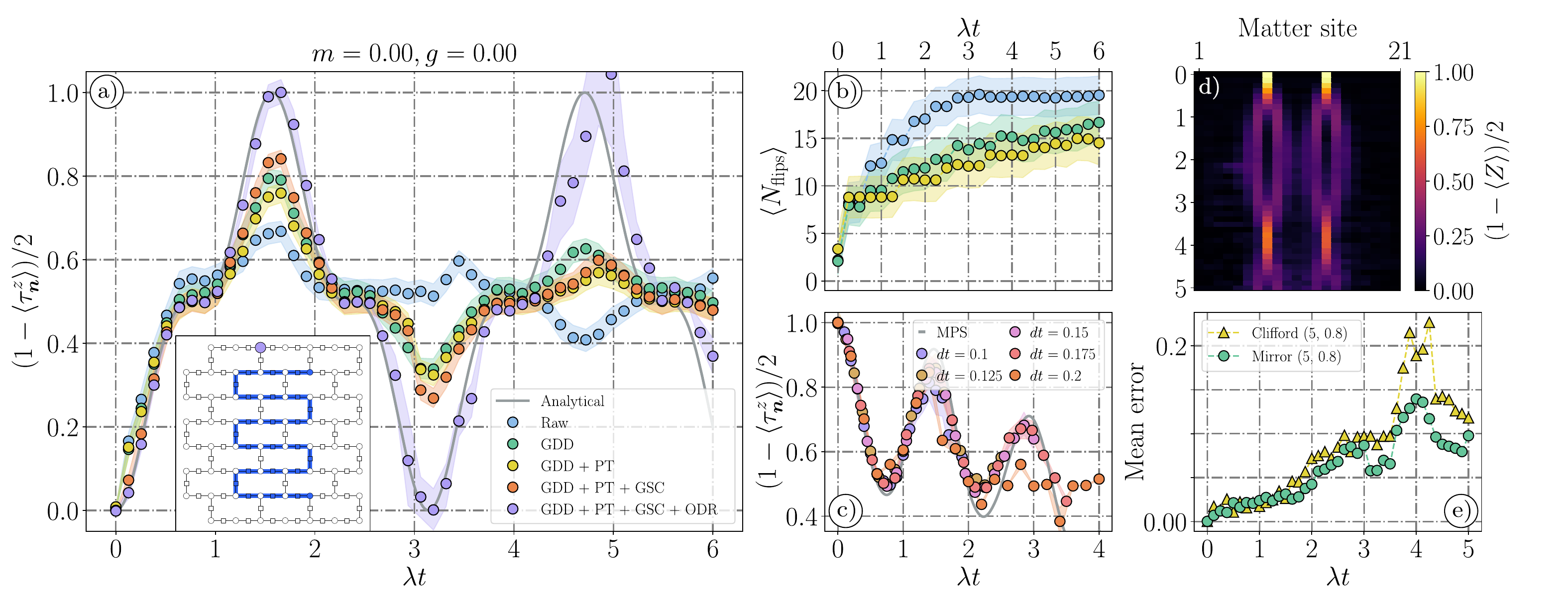}
\caption{\justifying Addressing the different sources of error in the simulation. (a) displays data for the dynamics of the local occupation in a trivalent matter site measured in the completely glassy regime $m=0$, $g=0$ in a $7\times 3$ lattice (144 qubits) for different error mitigation settings. Even though these dynamics can be reproduced with one Trotter layer, we choose $dt=0.25$ (7872 two-qubit gates) to evaluate the performance of the device with increasing depth. We progressively increase the number of error cancellation techniques introduced in the simulations and observe good convergence to the analytical expression. In (b), we quantify the number of flips measured by the GSC decoder for the different settings in (a). Dots represent the mean of the flip count distribution for the 300000 shots, and the shaded region indicates one standard deviation. (c) shows the occupation in the upper initial string endpoint from quantum simulations with varying $dt$ in the confined regime $m = 5$, $g = 2$, in the $2\times 2$ lattice (35 qubits). Every simulation reproduces the MPS results up to $dt=0.2$. In (d) we display the results of a quantum simulation of the $Z_2$HM in a (1+1)-dimensional chain of length $L=21$ (41 qubits) for $m=5$, $g=0.8$. The smaller dimensionality of the system constrains the propagation of errors, which allows performing a fair comparison of the mirror and Clifford ODR calibration circuit performance displayed in (e). We empirically observe that both calibration circuits perform similarly for short depth, but the mirror circuits are superior for long times, when more errors have accumulated, as Cliffordized circuits result in an effective change of the Hamiltonian coupling constants in the calibration.}
\label{fig:error_mitigation_data}
\end{figure*}

Leveraging the tunability of IBM devices, we can quench different initial states to explore novel dynamical effects such as a multi-string recombination. We consider a quench setting starting from the initial state of 3-strings colored blue in the legend of Fig.~\ref{fig:double_string_dynamics}(a), and evolving in the confined phase with parameters $m = 5\lambda$, $g = 2\lambda$. In this regime, the charges reduce their movements to nearby sites, which can result in the broken string configurations shown in Fig.~\ref{fig:double_string_dynamics} after first-order transitions from the initial state. This setting is not directly related to the usual string-breaking phenomenon for analogues of hadronization, recently observed in quantum devices~\cite{cochran2024visualizing,gonzalez2025observation}.

To probe multi-string recombination, we measure the following string-like correlators,
\begin{equation}
S_k = \prod_{n\in\mathcal{S}_k} \left( \frac{1- Z_n}{2} \right),
\end{equation}
where $\mathcal{S}_k$ is a set containing the indices of matter qubits that are the endpoints of the strings in each of the different configurations shown in Fig.~\ref{fig:double_string_dynamics}. For all configurations $|\mathcal{S}_k| = 4$. We choose these operators as probes of the multi-string configurations because if the state of the system coincides with one of these configurations, $\left\langle S_k \right\rangle = 1$. Roughly speaking, the configurations of Fig.~\ref{fig:double_string_dynamics} are not the only ones that lead to the expectation value $\left\langle S_k \right\rangle = 1$. However, these other configurations have either different particle numbers or differ in the field configuration, i.e., are separated from those we are probing by at least sixth-order perturbative processes; thus, we expect their contribution to be small. Measuring a four-body expectation value instead of a projector enables us to recover data from the quantum simulator without a large overhead in sample complexity due to noise. 

We can interpret the dynamics of Fig.~\ref{fig:double_string_dynamics} by considering leading-order transitions. The sharp decline observed in the population of the initial configuration, colored blue, corresponds to the initial tunneling of all of the matter particles to their neighboring sites. Notice that the first dip in the occupation occurs for $t = T_\mathrm{y}/2 \simeq 0.79$ because the process is controlled by first-order transitions as in the yo-yo mode. In this process, the broken-string configurations acquire a small population from the combined effect of partial tunneling of the matter particle in the center of the lattice and the partial non-tunneling of the rest. We thus refer to the matter particles not initially placed in the center as spectators. The small population acquired by the broken string configurations for short times decreases again when all these spectators completely tunnel out from their original positions. In the second yo-yo-like oscillation, the original matter sites start to become populated again, and so does the initial configuration. However, the central matter particle has partially drifted from its original position due to second-order effects, while the interior endpoints of the broken strings acquire population again from the yo-yo oscillations. During the same time, the spectators almost completely return to their original positions, and the broken string configurations experience a small peak in population with an amplitude bounded by the remnant of the central particle on the internal broken string endpoints. In Fig.~\ref{fig:supp_3string} of the extended data, we provide data for the local occupations in relevant matter sites that support this discussion. The results from this simulation demonstrate that the integration of error suppression, mitigation, and correction (see Sect.~\ref{subsec:methods_em}) enables the qualitative resolution of multi-string dynamics through nonlocal, four-body observables which take small expectation values on the order of $10^{-1}$.

\section{Methods} \label{sec:methods}

\subsection{Quantum Error Correction, Suppression \& Mitigation} \label{subsec:methods_em}

Current quantum simulators suffer from errors due to their interaction with the environment and imperfections in the pulse protocols that realize quantum gates. Although the long-term objective is the implementation of fault-tolerant algorithms, it is still possible to extract useful information from noisy simulations by introducing techniques that reduce the negative effects of noise without the large overhead of fully fault-tolerant approaches. Assuming that characterization and calibration can provide some knowledge of the noise channel, one can design various procedures to counteract it up to a certain error level~\cite{Cai_2023}. To perform the experiments in this work, we have developed two novel strategies: gauge dynamical decoupling (GDD) and Gauss sector correction (GSC), which are detailed in this section. In addition, we also use well-known strategies, namely Pauli twirling (PT) \cite{geller2013efficient} \cite{van2022model} and operator decoherence renormalization (ODR) \cite{farrell2024scalable}. We note that the latter can mitigate any Pauli noise channel when measuring tensor products of Paulis. We classify such procedures into three categories: correction, suppression, and mitigation. GSC is a correcting strategy that can perfectly correct for at least some kinds and numbers of errors. Error suppression (GDD, PT) refers to the modification of the circuits executed to actively reduce noise without altering the result of the simulations. ODR is an error mitigation technique that assumes a particular error channel and, after calibration of its parameters, extrapolates expectation values to the zero noise limit.

\subsubsection{Gauge dynamical decoupling} \label{subsubsec:methods_em_gdd}

Gauge dynamical decoupling (GDD) is an error suppression strategy intended to induce a random phase in non-physical states that may appear due to noise. It involves incorporating the gauge operators $G_{\boldsymbol{n}}$ into the generator of the time evolution operator so that the simulation circuits implement
\begin{equation}
\tilde{U}(t) = \prod_{k = 1}^L U_1(dt/2) \, U_3(dt) \, U_G(\left\{ \phi_{k, \boldsymbol{n}} \right\}) \, U_1(dt/2),
\end{equation}
instead of the usual Trotter decomposition of Eq.~\eqref{eq:second_order_trotter}, where we have defined
\begin{equation}
U_G(\left\{ \phi_{k, \boldsymbol{n}} \right\}) = \prod_{\boldsymbol{n}} {\rm e}^{-{\rm i} \phi_{k, \boldsymbol{n}}} G_{\boldsymbol{n}} ,
\end{equation}
with $\phi_{k, \boldsymbol{n}} \sim \mathrm{Uniform}[-\pi, \pi]$ random phases that fulfill the additional constraint $\sum_{k} \phi_{k, \boldsymbol{n}} = 0$. Since all the terms of Hamiltonian \eqref{eq:hamiltonian} commute individually with all the gauge transformation operators $G_{\boldsymbol{n}}$, it is possible to implement $U_G(\left\{ \phi_{k, \boldsymbol{n}} \right\})$ in the circuits without any overhead in depth. To do this, we follow the same strategy as outlined in Sect.~\ref{subsec:methods_trotter}. The set of random phases $\left\{ \phi_{k,\boldsymbol{n}} \right\}$ for a GDD-Trotter circuit of depth $L$ contains $LN_{\rm s}$ elements, with $N_{\rm s}$ the number of matter qubits, which coincides with the number of vertex stabilizer generators $G_{\boldsymbol{n}}$. The constraint $\sum_{k} \phi_{k, \boldsymbol{n}} = 0$ is interpreted as forcing the mean of random times associated with a single matter qubit to vanish in order not to introduce an unwanted bias. The set of random times is generated individually for each time instant of the simulations.

Gauge generators in the time evolution operator induce random phases in non-physical states, which helps average out the effects of errors. Effectively, one refocuses the dynamical events that push the system out of the physical Gauss sector back into the physical subspace. As a reminder, we call non-physical states those \textit{not} fulfilling Eq.~\eqref{eq:physical_states}. In general, we can write the density matrix at some particular instant of the noisy Trotter circuit as 
\begin{equation}
\rho(t) = (1-\eta(t))\,\rho_{\mathrm{p}}(t) + \eta(t)\,\rho_{\mathrm{np}}(t),
\end{equation}
where $\rho_{\mathrm{p}} \in \mathcal{H}_{\mathrm{p}}$, $\rho_{\mathrm{np}} \in \mathcal{H} / \mathcal{H_{\mathrm{p}}}$ and $\eta(t) \in [0, 1]$ is interpreted as an instantaneous parameter that quantifies the amount of unphysical states arising from environmental noise. From the commutation relation \eqref{eq:physical_states}, it follows that the Shrödinger conjugation of $U_G$ with $\rho(t)$ leaves the physical part of the density matrix invariant.
\begin{equation}
U_G \, \rho_{\mathrm{p}}(t) \, U_G^\dagger = \rho_{\mathrm{p}}(t).
\end{equation}
In contrast, it induces a random phase in the coherences of $\rho_{\mathrm{np}}(t)$,
\begin{equation}
\begin{aligned}
U_G \rho_{\mathrm{np}}(t) U_G^\dagger &= \sum_{a, b} c_{a, b} U_G \ket{a} \bra{b} U^\dagger_G \\
&= \sum_{a, b} c_{a, b} e^{-i \phi_{a, b}} \ket{a} \bra{b}.
\end{aligned}
\end{equation}
where the random phase $\phi_{a, b}$ follows the distribution,
\begin{equation}
\begin{aligned}
\phi_{a,b} \sim (&1-\delta_{a, b}) \\
& \times \left( \mathrm{Uniform}\left[ -\pi, \pi\right] + \mathrm{Uniform\left[-\pi, \pi\right]}\right).
\end{aligned}
\end{equation}

The introduction of this random phase causes the unphysical coherences to statistically vanish,
\begin{equation}
\mathbb{E}\left[ {\rm e}^{-{\rm i} \phi_{a, b}}\right] = 0,
\end{equation}
while keeping the true dynamics of the model unchanged. This drives $\rho_{\mathrm{np}}$ into a diagonal matrix, which implies that GDD contributes to the noise channel in the device being closer to a Pauli channel, increasing the effectiveness of ODR (Sect.~\ref {subsubsec:methods_em_odr}).

\subsubsection{Gauss sector correction}\label{subsubsec:methods_em_gse}

Let us now discuss how certain aspects of quantum error correction (QEC) can be exploited to correct certain hardware errors after measurement. We develop the novel Gauss sector correction (GSC) technique, which can formally correct single $X$-flips and also minimize higher-order flips by identifying them with very high probability. We start by noticing that the gauge constraint \eqref{eq:physical_states} defines a set $\{ G_{\boldsymbol{n}} \}$ of stabilizer and involutory operators $G_{\boldsymbol{n}}^2=\mathbb{I}$. When a bit-flip error occurs during the circuit, some of the gauge constraints are violated, and the state becomes unphysical. Some of these events are correctable using active QEC,  which will identify the most likely error and the recovery operation to revert it, and project back to the physical space. For an error to be recoverable, it must have a nonzero commutator with at least one of the stabilizer generators. We thus have a $G_{\boldsymbol{n}}$ stabilizer code such that bit-flips in any qubit (matter or gauge) are correctable, but phase flips remain completely undetected. Therefore, this stabilizer code cannot be used to achieve a full fault-tolerant procedure, but it can still provide some active error cancellation.

The bit-flip distance of our $G_{\boldsymbol{n}}$ code corresponds to $d = 3$. The physical subspace of our $Z_2$HM in Eq.~\eqref{eq:hamiltonian} is the linear span of all the possible products of individual interaction operators, plus the identity acting on the state $\ket{0000\dots0}$. The action of a single interaction operator is to create or destroy a pair of charges with minimal length, e.g., $\tau_{2}^x\sigma_{(1, 2)}^x \tau_{1}^x \ket{0000\dots0} = \ket{1110\dots0}$. This implies that valid codewords have a minimum Hamming distance of $d = 3$, which corresponds to the number of qubits affected by a single gauge-invariant interaction operator. This implies that formally, only $t=(d-1)/2=1$ bit-flip $X$-type errors are guaranteed to be fully correctable. However, if the bit flips occur in sufficiently distant qubits, the QEC protocol will correctly identify the error with very high probability and propose the correct recovery operation beyond $t=1$. The distance of the error-correcting code is defined as the minimum Hamming distance between any two valid codewords. 

We note that standard QEC implementations interleave gates from the original algorithm with so-called syndrome extraction rounds, where many-body stabilizer operators are mapped to ancillary qubits that can be measured without collapsing the logical state to detect errors \cite{548464}. Due to the nature of Trotter circuits, syndrome extraction could be performed by replacing the GDD matter qubit rotations with measurements in the $Z$ basis, without interfering with the dynamics of the system. However, in general, mid-circuit measurements can negatively affect the circuit performance due to the increased execution times and the extra errors induced by the measurement gadgets. For fault-tolerant protocols, there is a threshold error rate below which performing such a syndrome extraction provides an advantage despite the increased overhead, and once the physical error rates are below it, it is advantageous to scale up the QEC codes to gain protection and reduce logical error rates exponentially. This is not the case for our $G_{\boldsymbol{n}}$ code: As it cannot detect all types of errors, measuring the stabilizers in every Trotter layer causes more errors to appear than those that can be corrected. However, a final syndrome extraction round can always be performed without additional qubits and circuit overhead at the end of the circuit, since all matter and gauge qubits are measured in the $Z$ basis to infer the relevant physical observables. 

By multiplying the measurement results of the qubits in the support of each $G_{\boldsymbol{n}}$, we can determine if the associated gauge constraint is satisfied, and thus obtain the error syndrome. With this information, a QEC decoder can provide a recovery operation that corrects unphysical states. Since syndrome extraction is performed at the end of the quantum circuit, we are effectively performing classical error correction, and the decoder just indicates the measurement outcomes that it predicts to be flipped. We choose to use a minimum-weight matching decoder for this step. In particular, we use the \texttt{PyMatching2} implementation \cite{higgott2023sparse}, which offers remarkable performance with a C++ backend, together with a Python interface.

As we indicated previously, using the $G_{\boldsymbol{n}}$ code provided by the gauge constraints is not enough to achieve a fault-tolerant quantum simulation, which would require reencoding physical qubits in a different QEC code capable of correcting arbitrary errors, including also $Z$-type  phase flips. Thus, in the following lines, we provide a summary of all possible single-qubit errors that can occur during the circuit execution, and whether they can be corrected or otherwise impact the measured observables. We find the following effects by using standard propagation of errors~\cite{PhysRevA.99.062337} across rotations and CNOTs:

(i) When a single bit-flip $X$ error traverses an interaction layer, it remains unaffected, as it commutes with $H_I$. In principle, such bit-flips can be corrected by our stabilizer $G_{\boldsymbol{n}}$ code.

(ii) Bit-flip errors that occur within the interaction layer can propagate through the CNOT gates in the circuit, resulting in weight-2 and weight-3 errors. These higher-weight errors reflect the non-fault-tolerant nature of the circuit and fall outside the correction capabilities of the stabilizer code. These uncorrectable bit flips induce a single over/under-rotation in the interaction operator in which they happen, but future applications of the interaction operator remain unaffected.

(iii) When a single bit-flip $X$ error traverses a mass or electric field layer, it induces over/under rotations around the $Z$ axis, changing the mass or electric field of the qubit. Future applications of the operator are also affected. Bit flips by themselves are correctable, but the induced over-rotations, which are not detected by the QEC code, can affect the accuracy of the quantum simulation.

(iv) When a single phase-flip $Z$ error occurs and subsequently traverses an interaction layer implementing the evolution under $H_I$, or if it indeed occurs inside the interaction layer, it contributes with a gauge-invariant over/under-rotation generated by the operators $\tau_{\boldsymbol{n}+\boldsymbol{v}}^x\sigma_{(\boldsymbol{n},\boldsymbol{v})}^x\tau_{\boldsymbol{n}}^x$  that involve the faulty qubit.  Consequently, a single-phase flip error is translated into a coupling operator that will not be detected, suppressed, or mitigated by GDD and ODR. Moreover, future applications of the interaction operator will also be affected. Even though phase flips by themselves are, in principle, harmless, as they do not affect the final observables when occurring after the circuit, their propagation in the circuit makes them undetectable by the QEC code, and will eventually affect the accuracy of the quantum simulation.

(v) Measurement errors in the final step of the algorithm are equivalent to bit flips that occur just before the measurement. Thus, they are correctable by the QEC code. Since errors occurring in other parts of the circuit can affect dynamics without being detected, measurement errors are the main error source that is fully correctable by our GSC method.

In the experimental results presented in Sect.~\ref{sec:results}, we apply QEC by prioritizing the number of final samples used to estimate the observables instead of post-selecting the circuit runs that are strictly consistent with the physical Gauss sector. After retrieving the samples from the execution in quantum hardware, we sort them out by the number of flips detected by the decoder. We then set a minimum threshold of $30000$ samples and retain those with the fewest detected flips that meet this requirement. Fig.~\ref{fig:error_mitigation_data}(b) displays the average number of detected flips per circuit repetition as a function of circuit depth, for the benchmark simulations used to evaluate various error mitigation strategies. We observe that the number of flips increases rapidly at short-circuit depths and then transitions to a logarithmic growth at larger depths. This indicates that measurement errors are the dominant source of detected flips.

\subsubsection{Calibrating Operator Decoherence Renormalization} \label{subsubsec:methods_em_odr}

Operator decoherence renormalization (ODR) \cite{ciavarella2024quantum} is an error mitigation technique designed to eliminate the effect of Pauli noise on expectation values measured in noisy quantum hardware. Fig.~\ref{fig:error_mitigation_data}(a) shows that, in combination with Pauli Twirling \cite{geller2013efficient} and GDD (see Sect.~\ref{subsubsec:methods_em_gdd}), ODR is the most effective technique to mitigate errors in the expectation values of Pauli observables. ODR requires running an extra calibration circuit for each simulation circuit to estimate the combined probability of Pauli errors from hardware measurements. This effectively doubles the number of shots required. However, this is the minimum overhead found in any error mitigation technique that requires calibration. In this section, we describe the ODR and discuss the possible options for calibration circuits.

A Pauli quantum channel acting on a state $\rho$ of $N$ qubits is defined in general by the following set of Kraus operators,
\begin{equation}
\begin{aligned}
K_0 &= \sqrt{1  -\sum_i w_i} \ I, \\
K_i &= \sqrt{w_i} \bigotimes_{\boldsymbol{n}}\sigma^{\alpha_{i,\boldsymbol{n}}},
\end{aligned}
\end{equation}
where $\alpha_{i, \boldsymbol{n}} \in \{0, x, y, z\}$ is an index that indicates whether the identity or a Pauli matrix is acting on the qubit $\boldsymbol{n}$ and we implicitly assume that for each $i$, at least one $\alpha_{i, \boldsymbol{n}} \neq 0$. The index $i = 1, 2, \dots 4^N - 1$ runs through all the possible Pauli errors that can occur with an associated error weight $w_i\in[0,1]$. This Pauli noise channel acts, by convention, after the $\mathcal{U}$ unitary gates included in the noiseless ideal circuit $\rho_\mathrm{f} = (\mathcal{D} \circ \mathcal{U})[\rho_0] = \mathcal{D}[\rho_\mathrm{u}]$, the noisy expectation value of an observable $O$ is
\begin{equation}
\langle O \rangle_{\mathrm{f}} = \left( 1 - \sum_{i} w_i \right) \langle O\rangle_{\mathrm{u}} + \sum_{i} \tr\left(K_i^{\dagger} \rho_\mathrm{u} K_i O \right).
\label{eq:odr_general}
\end{equation}

We are interested in observables $O$ that can be expressed as tensor products of Pauli operators, such that the following identity holds,
\begin{equation}
\tr \left(K_i^\dagger \rho_{\mathrm{u}} K_i O \right) = \pm w_i \tr \left( \rho_{\mathrm{u}} O \right).
\label{eq:k_i_term}
\end{equation}
Combining Eqs.~\eqref{eq:odr_general} and~\eqref{eq:k_i_term}, the noiseless and noisy expectation values are related linearly,
\begin{equation}
\langle O \rangle_{\mathrm{f}} = (1-p) \langle O\rangle_{\mathrm{u}},
\label{eq:odr_final}
\end{equation}
where we have defined $p = 2\sum_{i\in\mathcal{A}_O} w_i$ as the sum of weights of Kraus operators that anticommute with the specific observable $O$. Eq.~\eqref{eq:odr_final} implies that the noiseless expectation value $\langle O \rangle_\mathrm{u}$ can be obtained from $\left\langle O \right\rangle_{\mathrm{f}}$ provided one can estimate the value of $p$. This can be done using an extra calibration circuit with an efficiently computable ideal output $\langle O \rangle_\mathrm{u}$, such that $p$ can be solved from Eq.~\eqref{eq:odr_final} by dividing the measured noisy outcome and the ideal precomputed one. The key is that the calibration circuit must be as similar as possible to the ones used in the real quantum simulation, such that we can ensure the minimal distance between the two quantum channels $(\mathcal{D} \circ \mathcal{U})_{\mathrm{cal}}$ and $(\mathcal{D} \circ \mathcal{U})_{\mathrm{sim}}$. It is important to note that we apply the same setting for all other error suppression and correction strategies in the simulation and calibration circuits. The other key aspect is that the noise must be accurately described by Pauli noise, and this is the reason why the PT integrates nicely with ODR.

For calibration, we choose to use mirror circuits, also known as quantum Loschmidt echoes. Since Trotter circuits are symmetric for an even number of layers, we can design a calibration circuit by cutting the original simulation circuit in half, and inverting the second half in an effective time reversal, such that computing $O_u$ is trivial for an initial product state. The resulting pair of simulations and calibration circuits shares a very similar structure, yielding, to the best of our experience, the most precise calibration we have empirically found. An alternative for calibration is to use Cliffordized circuits \cite{Aaronson_2004}. These are generated from the original simulation circuit by substituting the angle of non-Clifford rotations with the closest Clifford angle. In Fig.~\ref{fig:error_mitigation_data}(f), we compare the relative error between $(1+1)$D quantum and MPS-based simulations achieved with the two calibration strategies. We reduce the dimensionality of the model to gain stability in the results for the comparison. We observe that mirror circuits achieve better precision for the longer times, as errors appearing in the second half of mirror circuits propagate exactly as in the simulation circuits, which is not the case for the Clifford circuits as a result of effective changes in the microscopic parameters.

\subsection{Ground state DMRG calculations} \label{subsec:methods_dmrg}

The ground state phase diagram of the model has been studied through large-scale DMRG simulations~\cite{White-1992, Schollwock-2011}. To avoid anisotropy effects, flower-like geometries have been chosen that reflect the six-fold rotational symmetry of the lattice. The size of the system is parameterized by a single parameter $R$, which is the number of layers of hexagons around the central hexagon: we have a single hexagon for $R=0$, $7$ hexagons for $R=1$, and $19$ hexagons for $R=2$ (see also Extended Data Fig.~\ref{fig:phasediag_numerics}(a)).  We form super-sites from tensor products of local Hilbert spaces, of one matter ($\tau$) and its three neighboring gauge ($\sigma$) spins. Projection to physical states, i.e., fixing the gauge in \eqref{eq:physical_states}, leads to $8$-dimensional local Hilbert spaces of super-sites. However, we note that supersite formation results in the duplication of gauge spins ($\sigma_{n,v} \Rightarrow \lbrace \tilde{\sigma}_{n,v}^{(1)}, \tilde{\sigma}_{n,v}^{(2)} \rbrace  $), and the corresponding spin pairs must be enforced to be parallel. Therefore, a term $H_U = U \sum_{(n,v)} \tilde{\sigma}_{(n,v)}^{z (1)} \tilde{\sigma}_{(n,v)}^{z (2)} $, which commutes with $H$, is formally added to \eqref{eq:hamiltonian} to ensure the removal of unwanted and unphysical antiparallel configurations. Within the super-site framework, the Hamiltonian \eqref{eq:hamiltonian} contains only 1-site and 2-site terms, but due to the two-dimensional geometry, some 2-site couplings get long-ranged. However, our flexible DMRG implementation can efficiently treat such situations~\cite{Menczer-2024b}. Within this work, the largest simulated system sizes are $R=2$, which corresponds to $N_n=54$ super-sites (nodes). The DMRG algorithm results in a matrix product state (MPS) wave function~\cite{Schollwock-2011}, whose precision is controlled by the so-called bond dimension. The largest bond dimension is set to $M=256$ for the $g-m$ mesh in Fig.~\ref{fig:fig0}(a) and Extended Data Fig.~\ref{fig:phasediag_numerics}(b-d), while for the cuts in Extended Data Fig.~\ref{fig:phasediag_numerics}(e-g), the bond dimension has been increased to $M=1024$. Note that in addition to the gauge symmetries already exploited, there are no further conserved quantum numbers; therefore, diagonalization of an effective Hamiltonian of dimension $M^2\times8^2$ is inevitable. The detailed accuracy thresholds and system sizes described above are enough to sketch the phase diagram qualitatively, while the precise determination of the phase boundaries requires further investigations beyond the current scope. 

In the extended data in Fig.~\ref{fig:phasediag_numerics}, numerical results are presented to supplement those in Fig.~\ref{fig:fig0}.  The ground-state magnetizations of matter and gauge spins, $\langle \tau^z \rangle$ and $\langle \sigma^z \rangle$, are shown in panels (b) and (c) for the regions $m \in [0,3]$ and $g \in [0,2]$. We observe a low magnetization region (for both $\tau^z$ and $\sigma^z$) in the bottom left corner, around the $m=0$, $g=0$ point. For large values of both $m$ and $g$, the magnetizations are close to $1$. However, for $m\gtrsim 1.5$ and $g \ll 1$ only the matter spins are fully polarized, while the magnetization of the gauge spins is $\langle \sigma^z \rangle < 1$. We identified the low-magnetization region with the Higgs regime, the fully polarized region with the confined regime, and the narrow region around the horizontal axis with the deconfined phase.

A similar behavior is found by calculating the entanglement (von Neumann) entropy between two halves of the system. Here, our flower-like geometry is cut in two symmetrically by a vertical line, and the entropy is calculated from the so-called Schmidt values that are already available in the DMRG algorithm~\cite{Legeza-2003}. We observe a high entropy in the Higgs regime, a low entropy in the confined regime, and an elevated entropy in the narrow deconfined region. Interestingly, the bond dimension $M=256$ is enough for the high entropy state in the Higgs regime, but insufficient in the narrow deconfined region (see also Extended Data Fig.~\ref{fig:phasediag_numerics}(e-g)). The extension of the deconfined phase can be approximated by $\tilde{g} \approx J_{\hexagon}^{\mathrm{eff}} = \gamma \lambda^6 m^{-5}$ with $\gamma\approx 0.25$. This approximate phase boundary is also indicated by vertical dashed lines in panels (e-g) and found to coincide with the edge of the elevated entropy. Similar results have also been obtained for smaller systems with $R=1$ and $R=0$. 

\section{Conclusions and Outlook}

We have demonstrated the real-time quantum simulation of string dynamics in a (2+1)-D LGT using a superconducting quantum chip. By encoding the full gauge-matter dynamics of the $Z_2$HM, we observed coherent phenomena including string oscillations, endpoint-induced bending, and multi-string fragmentation,
also validated by extensive tensor network simulations. These effects constitute a direct manifestation of the emergent string-like nature in engineered gauge theories. Our work represents a critical advance in the quantum simulation of non-equilibrium LGTs, bridging the gap between theoretical constructs and experimental observables. Our implementation illustrates how programmable quantum systems, despite hardware limitations, can access rich real-time behavior when pushing the limits of error mitigation, suppression, and correction. 

Looking ahead, several directions emerge. First, scaling to larger system sizes and increasing gate fidelity to allow for larger circuit depths will permit the exploration of other collective string phenomena, such as string-string scattering or the existence of bound states. Second, implementing more general gauge groups (e.g., $Z_n$, $U(1)$) would extend the reach of these methods toward other richer symmetry structures, where the interplay with anomalies and the dynamical matter could be explored. Third, integrating variational state preparation via error-robust algorithms~\cite{cobos2024noise} may unlock access to controlled quenches from ground or thermal states, allowing dynamical probes of confinement, string breaking, and phase transitions. Ultimately, the intersection of quantum information science and LGTs, extending the present connections to QEC towards full fault-tolerant approaches, offers a unique opportunity to experimentally probe the real-time quantum dynamics of extended objects, ushering in a new era for string physics and non-perturbative quantum field theory.

\section{Acknowledgments}

The authors acknowledge discussions and input on the project from K. Temme; they also acknowledge fruitful discussions on the formalism of quantum and classical error correction codes with J. Etxezarreta Martinez at IBM's QDC 2024.  We acknowledge IBM’s Quantum Algorithm Engineering team for their insights and contributions.

We acknowledge BasQ-Ikerbasque for the access to the IBM machines that allow us to complete this project. We thank the discussions in the QC4HEP working group.

E.R. acknowledges the financial support received from the IKUR Strategy under the collaboration agreement between the Ikerbasque Foundation and UPV/EHU on behalf of the Department of Education of the Basque Government. J.C. F.d.M. and E.R. acknowledge support from the BasQ strategy of the Department of Science, Universities, and Innovation of the Basque Government. E.R. is supported by the grant PID2021-126273NB-I00 funded by MCIN/AEI/10.13039/501100011033 and by “ERDF A way of making Europe” and the Basque Government through Grant No. IT1470-22. This work was supported by the EU via QuantERA project T-NiSQ grant PCI2022-132984 funded by MCIN/AEI/10.13039/501100011033 and by the European Union “NextGenerationEU”/PRTR. This work has been financially supported by the Ministry of Economic Affairs and Digital Transformation of the Spanish Government through the QUANTUM ENIA project, called Quantum Spain project, and by the European Union through the Recovery, Transformation, and Resilience Plan – NextGenerationEU within the framework of the Digital Spain 2026 Agenda.

This work has been partially funded by the Eric \& Wendy Schmidt Fund for Strategic Innovation through the CERN Next Generation Triggers project under grant agreement number SIF-2023-004.

This work has also been supported by the Hungarian National Research, Development and Innovation Office (NKFIH) through Grant Nos.~K134983 and TKP2021-NVA-04, and by the Quantum Information National Laboratory of Hungary. \"O.L. acknowledges financial support by the Hans Fischer Senior Fellowship programme funded by the Technical University of Munich – Institute for Advanced Study and from the Center for Scalable and Predictive methods for Excitation and Correlated phenomena (SPEC), funded as part of the Computational Chemical Sciences Program FWP 70942 by the U.S. Department of Energy (DOE), Office of Science, Office of Basic Energy Sciences, Division of Chemical Sciences, Geosciences, and Biosciences at Pacific Northwest National Laboratory.

M.A.W. has also been supported by the Janos Bolyai Research Scholarship of the Hungarian Academy of Sciences.

A.B. and C.B. acknowledge support from PID2021-127726NB- I00 (MCIU/AEI/FEDER, UE), from the Grant IFT Centro de Excelencia Severo Ochoa CEX2020-001007-S, funded by MCIN/AEI/10.13039/501100011033, from the CSIC Research Platform on Quantum Technologies PTI-001.

JC is supported by the PIF 2022 grant funded by UPV/EHU.

\newpage

\bibliography{references.bib}

\clearpage

\appendix

\onecolumngrid

\section{Supplemental material}

\subsection{Solution for the $Z_2$HM in the $m = 0$ limit} \label{supp:m_0_analytics}

In case of $m = 0$, the Hamiltonian reads 
\begin{equation}
H =  \sum_{(\boldsymbol{n},  \boldsymbol{v})} h_{(\boldsymbol{n},  \boldsymbol{v})} = \sum_{(\boldsymbol{n},  \boldsymbol{v})}  \left( - g  \sigma_{(\boldsymbol{n},  \boldsymbol{v})}^{z} - \lambda  \tau_{\boldsymbol{n} + \boldsymbol{v}}^x \sigma_{(\boldsymbol{n}, \boldsymbol{v})}^x \tau_{\boldsymbol{n}}^x \right) \; .
\label{eq:mzero_hamiltonian}
\end{equation}
Here, the sum goes over every bond $(\boldsymbol{n},  \boldsymbol{v})$, and the single-bond Hamiltonians $h_{(\boldsymbol{n},  \boldsymbol{v})} =  - g  \sigma_{(\boldsymbol{n},  \boldsymbol{v})}^{z} - \lambda  \tau_{\boldsymbol{n} + \boldsymbol{v}}^x \sigma_{(\boldsymbol{n}, \boldsymbol{v})}^x \tau_{\boldsymbol{n}}^x $ commute with each other, $[h_{(\boldsymbol{n},  \boldsymbol{v})}, h_{(\tilde{\boldsymbol{n}},  \tilde{\boldsymbol{v}})}] \equiv 0$.

\subsubsection{Ground state solution}

To find the ground state, first, we ignore the gauge fixing. The single-bond Hamiltonian $h_{(\boldsymbol{n},  \boldsymbol{v})}$ is an $8 \times 8$ matrix that becomes immediately block diagonal if we express it in the $x$-basis: if we fix the two matter spins in parallel configuration in the $x$-basis, i.e., $(\leftarrow\leftarrow)$ or $(\rightarrow \rightarrow)$, then the gauge spin's state is given by the eigenstate of the  $2 \times 2$ sized block $M_{\mathrm{FM}}$, while for antiparallel matter spins $(\leftarrow\rightarrow)$ or $(\rightarrow \leftarrow)$ the matrix block $M_{\mathrm{AFM}}$ has to be diagonalized, where
\begin{equation}
M_{\mathrm{FM}} = \left( \begin{array}{cc} -\lambda & -g \\-g & +\lambda \end{array} \right) \, , \; \mathrm{and} \quad M_{\mathrm{AFM}} = \left( \begin{array}{cc} +\lambda & -g \\-g & -\lambda \end{array} \right) \; .
\end{equation}
The eigenvalues of $M_{\mathrm{FM}}$ and $ M_{\mathrm{AFM}}$ are $\pm \sqrt{\lambda^2 + g^2}$. Consequently, independently of the configuration of the matter spins, we can minimize the energy of the bond, and this minimum is $-\sqrt{\lambda^2 + g^2}$.

The ground state of \eqref{eq:mzero_hamiltonian} is highly degenerate if the gauge fixing is ignored: One can freely choose a random configuration of the matter spins in the $x$-basis, and then the state of the gauge spins is set one by one simply to the ground state of $M_{\mathrm{FM}}$ or $M_{\mathrm{AFM}}$ depending on the configuration of the neighboring matter spins. Without gauge constraint, the ground state subspace is therefore $2^{N_n}$ times degenerate. However, this degeneracy is removed when the gauge sector is fixed. Here we note that, if we follow the construction above, the resulting state will have nonzero overlap with every product basis state in the $z$-basis, in which basis the gauge operators are diagonal. Consequently, every ground state of our construction has a non-zero overlap with every gauge sector. Since there are $2^{N_n}$ different gauge sectors, there will be exactly one ground state in every gauge sector, which is obtained after projecting the unconstrained state into the selected subspace. Following this reasoning, and noticing that the ground state of the model has a simple structure when $g\to0$ or $\lambda \to 0$, we design a state ansatz providing the exact expression for the ground state in the physical sector $G_{\boldsymbol{n}}\ket{\psi} = \ket{\psi}$. The ansatz is the following,
\begin{equation}
\ket{\psi_0(g, \lambda)} = \left\{\cos \theta(g, \lambda) \ \mathbb{I} + \sin \theta(g, \lambda) \left[\prod_{(\boldsymbol{n}, \boldsymbol{v})} \left( \frac{\mathbb{I} + \tau_{\boldsymbol{n} + \boldsymbol{v}}^x \sigma_{(\boldsymbol{n}, \boldsymbol{v})}^x \tau_{\boldsymbol{n}}^x}{\sqrt{2}} \right) - \mathbb{I} \right] \right\} \ket{000\dots0},
\label{eq:ansatz_m0}
\end{equation}
which interpolates between the $\ket{\psi_0(g,0)} = \ket{000\dots0}$ state and an equal superposition of all the states in the physical basis,
\begin{equation}
\ket{\psi_0(0, \lambda)} = \prod_{(\boldsymbol{n}, \boldsymbol{v})} \left( \frac{\mathbb{I} + \tau_{\boldsymbol{n} + \boldsymbol{v}}^x \sigma_{(\boldsymbol{n}, \boldsymbol{v})}^x \tau_{\boldsymbol{n}}^x}{\sqrt{2}} \right) \ket{000\dots0},
\end{equation}
which are the ground states of Hamiltonian \eqref{eq:mzero_hamiltonian} in the regimes $g = 0$ and $\lambda = 0$ respectively. The value of the variational parameter is obtained analytically through the variational principle,
\begin{equation}
\theta(g, \lambda) = \frac{1}{2} \arctan \left( \frac{\lambda}{g} \right).
\end{equation}

This ansatz can be interpreted as a product of gauge-invariant rotations in each lattice bond and interpolates between a product state and a long-range entangled state. It ensures that the energy associated with the state of the gauge qubits is minimal for every $g$, $\lambda$ while ensuring that the gauge constraint is fulfilled. Since there is always a finite gap, this interpolation can always yield the exact ground state. 

The gap of Hamiltonian \eqref{eq:hamiltonian_m0} can be computed following a similar reasoning as for the ground state. In the gauge-unrestricted construction, excited states are those in which one gauge spin is flipped from its ground state to the excited state. The gap is then $\Delta E = 2 \sqrt{\lambda^2 + g^2}$. Again, every excited state constructed in this way will have a nonzero overlap with every gauge sector, i.e., gauge fixing does not affect the value of the gap. One can check that the ansatz \eqref{eq:ansatz_m0} also reproduces the value of the gap. Note that Hamiltonian \eqref{eq:hamiltonian_m0} contains two terms that correspond to two distinct gauge-invariant excitations. The crossover between fundamental excitations occurs at $g = \lambda$, which coincides with the point at which $\langle \sigma_{(\boldsymbol{n}, \boldsymbol{v})}^z \rangle \simeq 0.5$ in Fig.~\ref{fig:phasediag_numerics}.

\subsubsection{Dynamics}

The structure of Hamiltonian \eqref{eq:hamiltonian_m0}, which is a sum of commuting operators, leads to a simple expression for the dynamics of expectation values in the quench setting. In the Heisenberg picture, observables $O(t)$ are the dynamical variables, evolving as
\begin{equation}
O(t) = U^\dagger(t) \, O(0) \, U(t).
\end{equation}
with $U(t)$ the time evolution operator. For the $\tau_{\boldsymbol{n}}^z$ operators living on sites connected to $d = \{2, 3\}$ bonds, $U(t)^\dagger \tau_{\boldsymbol{n}}^z U(t)$ is a sum of $3^d$ terms appearing from the different combinations of left and right products of the identity and the different gauge-invariant excitations $\{ \mathbb{I} \} \cup \{\sigma_{(\boldsymbol{n}, \boldsymbol{v})}, \tau_{\boldsymbol{n} + \boldsymbol{v}}^x \sigma_{(\boldsymbol{n}, \boldsymbol{v})}^x \tau_{\boldsymbol{n}}^x \}_{\boldsymbol{v}}$, with appropriate constants. For local gauge operators $\sigma_{(\boldsymbol{n}, \boldsymbol{v})}^z$, only one link term $h_{(\boldsymbol{n}, \boldsymbol{v})}$ contributes and $d=1$. Of these $3^d$ terms, there is only one with a non-vanishing expectation value when evaluated over the quench initial state, corresponding to the original observable, which leads to a concise expression for the quench dynamics
\begin{equation}
\langle O_d(t) \rangle = \left[ 1 - \frac{2\lambda^2}{\lambda^2 + g^2} \, \sin^2 \left( t\sqrt{\lambda^2 + g^2} \right) \right]^d \langle O_d(0) \rangle.
\end{equation}

\subsection{Trotter circuits} \label{subsec:methods_trotter}

To simulate the dynamics of $Z_2$HM on quantum hardware, we implement the time evolution operator by its second-order Trotter decomposition,
\begin{equation}
\begin{aligned}
U(t) &= \prod_{k = 1}^L U_1(dt/2) \, U_3(dt) \, U_1(dt/2) \\
& = \prod_{k = 1}^L e^{-i H_1 dt/2} \, e^{-i H_3 dt} \, e^{-i H_1 dt/2}
\end{aligned}
\label{eq:second_order_trotter}
\end{equation}
With $H_1 = -H_M - H_E$, $H_3 = -H_I$, the one and three local terms of Hamiltonian \eqref{eq:hamiltonian} respectively and $t = L\, dt$. Here, we have defined $L$ as the number of Trotter layers. Our choice of the second-order Trotter decomposition is motivated by the fact that it requires the same number of entangling gates as the first-order decomposition while reducing the approximation error with the real dynamics. Currently, quantum algorithms are primarily limited by the number of entangling gates in the circuit to be executed, making higher-order decompositions less practical, despite being more precise in theory.

The circuits implementing the second-order Trotter decomposition \eqref{eq:second_order_trotter} are shown in Fig.~\ref{fig:fig0}(d). The operator $U_1$ is a tensor product of single-qubit rotations; thus, its implementation is straightforward. $U_3$ is implemented as a composition of one Pauli gadget of Fig.~\ref{fig:fig0}(e) per lattice edge. This Pauli gadget includes single-qubit rotations to implement Gauge Dynamical Decoupling (see Sect.~\ref{subsec:methods_em}). To achieve the optimal entangling gate depth of 6 for the implementation of $U_3$, we divide the matter qubits in the lattice into two disjoint sets that we depict with the different colors of the CNOT gates in Fig.~\ref{fig:trotter_circuits} of the supplemental information. Notice that matter qubits are always the target of these CNOTs, and gauge qubits always act as controls. We define these disjoint sets to ensure that each gauge qubit only acts as control of a single CNOT in each layer of the circuit. To design the $U_3$ circuit, we placed two consecutive layers of Pauli gadgets, one per set of matter qubits. In each layer, we place one Pauli gadget per matter qubit in the set, with the CNOTs ordered as in Fig.~\ref{fig:trotter_circuits}. This strategy is designed such that all the CNOTs of the two layers of Pauli gadgets commute, so that everything can be combined into a single dense block. This strategy ensures the minimal depth of the resulting circuit and increases the gate density, reducing the proliferation of errors. The final simulation circuit is built from the sequential application of $U_1$ and $U_3$

\begin{figure*}[!ht]
\includegraphics[width=0.9\textwidth]{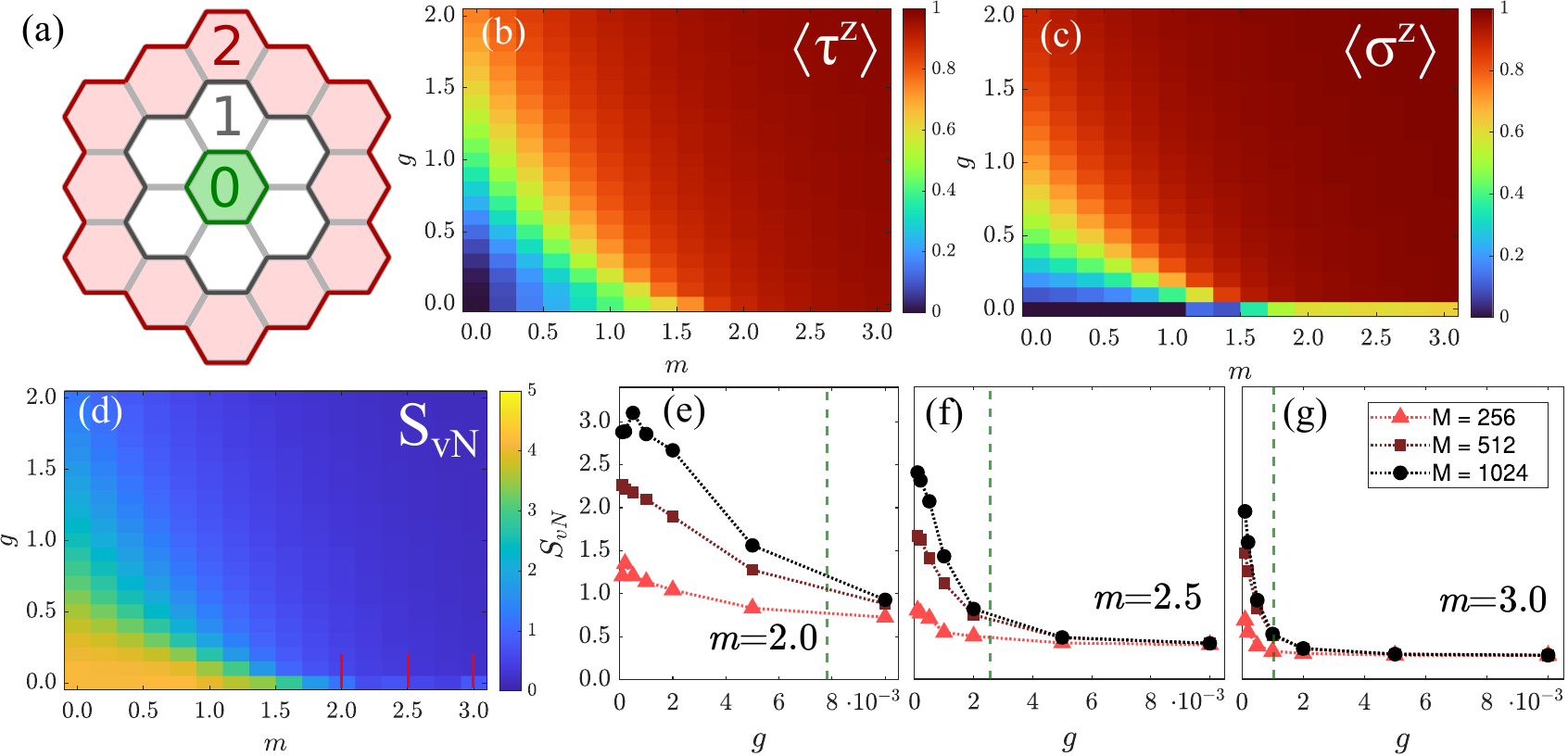}
\caption{\justifying DMRG results for the ground state. Panel (a) shows the simulated symmetric flower-like flakes. Numbers indicate system sizes for $R \in [0,1,2]$. The largest simulated system contains 19 hexagons and $N_n = 54$ nodes. In panels (b-c), the average magnetizations of the matter and gauge qubits are shown, respectively, for system size $R=2$ and bond dimension $M=256$. The bottom left corner with low magnetization corresponds to the Higgs region, the large red region with magnetization close to one indicates the confined regime. Close to the horizontal axis, the matter spins are fully polarized, but the gauge spins are not: this indicates a narrow deconfined region. In panel (d), the ground state entanglement (von Neumann) entropy is shown between the two system halves if the flake of size $R=2$ is cut symmetrically by a vertical line at the middle. The bond dimension is again $M=256$. Vertical red lines at the bottom right indicate the specific cuts of panels (e-g) for which higher accuracy (up to $M=1024$) DMRG calculations have been performed. In these panels, vertical dashed lines indicate the roughly proposed phase boundary $\tilde{g} \approx J_{\hexagon}^{\mathrm{eff}}$ between the deconfined and confined phases. We observe elevated entropy for $g< J_{\hexagon}^{\mathrm{eff}}$, and also the DMRG did not fully converge even for $M=1024$. In contrast, we observe low entropy and reasonable convergence for $g > J_{\hexagon}^{\mathrm{eff}}$.}
\label{fig:phasediag_numerics}
\end{figure*}

As expected from the theory of the Trotter error \cite{PhysRevX.11.011020}, we can only approximate the desired dynamics with finite depth. The tightest bound known for the additive error of the second-order Trotter decomposition $\varepsilon = \| U(t) - e^{-i H t} \|$ is
\begin{equation}
\varepsilon \leq \frac{t^3}{12} \| \left[ H_3, \left[ H_3, H_1\right] \right] \| + \frac{t^3}{24} \| \left[ H_1, \left[ H_1, H_3\right]\right] \|,
\label{eq:trotter_error_general}
\end{equation}
where $\left\| \cdot \right\|$ is the operator spectral norm. Putting in the explicit expression for $H_1$ and $H_3$, and after some algebra, the expression for the Trotter error becomes
\begin{equation}
\begin{aligned}
\varepsilon \leq & \frac{t \, dt^2}{12} \left[ 4N_{e} \left| g\lambda^2 \right| + \left( 16 N_{n,2} + 36 N_{n, 3} \right) \left| m\lambda^2 \right| \right] \\
& \begin{aligned}
+ \frac{t\, dt^2}{24} \left[ 8N_e\left| m^2 \lambda \right| \right. + & \left. 16N_e \left| mg\lambda \right|  \right. \\
& \left. + N_e \left| 4g^2\lambda + 8m^2\lambda \right| \right]
\end{aligned} \\
\end{aligned}
\label{eq:trotter_error_z2higgs}
\end{equation}
where we have defined $N_e$ as the number of edges in the lattice and $N_{n,2}$, $N_{n, 3}$ the number of nodes connected to two and three edges, respectively. Notice that the Trotter error scales linearly with the lattice size and with the different products of the three coupling constants $m, g, \lambda$ with distinct coefficients. Thus, Eq.~\eqref{eq:trotter_error_z2higgs} implies that, for fixed circuit depth, the Trotter error will be larger in the top and right regions of the phase diagram of Fig.~\ref{fig:phasediag_numerics}. Since depth is one of the hardest constraints for successfully executing a circuit in current quantum hardware, this justifies the fact that it is more difficult to reconstruct the dynamics of the model in the confined regime for quantum hardware.

After having experimented with different settings, we set a limit of $204$ CNOT depth for the simulation circuits. Although we certainly observe more noise in larger lattices, we empirically found that this is a reasonable threshold to obtain decent results. Considering the CNOT depth of 6 required to implement $U_3$, this amounts to a maximum depth of the Trotter layer of $L = 34$. In Tab.~\ref{tab:trotter_error}, we show the Trotter error predicted by bound \eqref{eq:trotter_error_z2higgs} for the different parameters that we simulate on quantum hardware. Notice that in all cases, except the diagonal $3\times7$, the error is predicted to take the value $\varepsilon>1$. This seems to indicate that the depth that we can currently reach is not sufficient to properly reconstruct the dynamics of the model in the most demanding regime. However, as the authors of \cite{PhysRevX.11.011020} indicates, there is a margin between the error predicted by the bound \eqref{eq:trotter_error_general} and the true error. To check that we are using the shortest depth possible, to minimize noise, while being able to properly reconstruct the dynamics, we have run various simulations for each of the parameters with different $dt$. Fig~\ref{fig:trotter_circuits}(c) shows the dynamics of the occupation in the upper initial string endpoint for $m = 5, g = 2$, in the $2 \times 2$ lattice for different $dt$. From this figure, we estimate that $dt \leq 0.175\lambda$ is enough to properly approximate the dynamics up to noise error. In the simulations presented in Sect.~\ref{sec:results}, we force the number of layers of the Trotter circuits $L$ to be even. The reason for this is the calibration step for the ODR error mitigation technique, which is described in Sect.~\ref{subsubsec:methods_em_odr}. When generating the circuits in Qiskit, we first choose $dt$ to fix the error, then, for each $t$, we first calculate the number of layers as $L = \left\lceil t/dt \right\rceil$, and if it is odd, we add one. In case $t/L$ is not an integer, each layer receives a smaller $d\tilde{t} = t/L$.

\begin{table}[]
\centering
\begin{tabularx}{0.8\textwidth}{ 
| >{\centering\arraybackslash}X |
| >{\centering\arraybackslash}X 
| >{\centering\arraybackslash}X 
| >{\centering\arraybackslash}X |
| >{\centering\arraybackslash}X 
| >{\centering\arraybackslash}X |
| >{\centering\arraybackslash}X |}
\hline
Lattice Size (Qubits) & $m$ & $g$ & $\lambda$ & $t$ & $dt$ & $\varepsilon$ bound \eqref{eq:trotter_error_z2higgs} \\ \hline \hline \vspace{1pt}
$2\times2$ (36) & $5$ & $2$ & $1$ & 4 & 0.15 & 50.28 \\[1pt] \hline \vspace{1pt}
$3\times3$ (68) & $5$ & $0.01$ & $1$ & 4 & 0.125 & 60.50 \\[1pt] \hline \vspace{1pt}
$3\times3$ (68) & $0.3$ & $0.5$ & $1$ & 4 & 0.15 & 4.31 \\[1pt] \hline
$7\times3$ (146) & $0$ & $0$ & $1$ & 6 & 0.25 & 0 \\[1pt] \hline
\end{tabularx}
\caption{\justifying Trotter error bound for the quantum simulations performed for the different phases of the model. The depths accessible to the quantum device are not sufficient to formally guarantee precision in every regime, as $\varepsilon > 1$; however, we do not observe variation when increasing $dt$ (Fig.~\ref{fig:error_mitigation_data}). Bound \eqref{eq:trotter_error_z2higgs} can be distant from the true Trotter error in general, yet it is the tightest bound currently known \cite{PhysRevX.11.011020}.}
\label{tab:trotter_error}
\end{table}

\subsection{MPS-based dynamical simulations} \label{supp:methods_mps_dyn}

MPS wavefunctions provide a framework not only for computing the eigenvalue spectrum of static Hamiltonians but also for performing simulations of quantum dynamics. As the two-dimensional geometry introduces long-range couplings in the MPS chain, it is beneficial to use a time integrator algorithm that, similarly to our DMRG implementation, can treat these straightforwardly~\cite{Haegeman-2016, Ceruti-2023}. Due to the large, 8-dimensional, supersite of our MPS, we chose the recently introduced ``basis update and Galerkin" (BUG) integrator~\cite{Ceruti-2023}, which uses a one-site scheme but with adaptive bond expansion. The wavefunction has always been initialized in a simple product state that can be generated from a spin-polarized state by flipping some gauge and matter qubits. Importantly, the initial states in our simulations are always physical, i.e, $G_n \ket{\psi} = \vert\psi \rangle  ~\forall n$. Magnetizations $\langle \sigma_{(m,v)}^z(t) \rangle$, $\langle \tau_{n}^z(t) \rangle$, and various equal time correlations have been determined from the time-dependent wavefunction that are directly comparable with results of quantum hardware. The bond dimension for these simulations has been varied between $D \in [64-512]$ depending on the parameters $m$ and $g$, and convergence has been ensured by comparing results for different bond dimensions. We note one important aspect of our dynamical simulations, namely, the initial bond expansion. The initial state $\ket{\psi_0}$ is always a product with an MPS of bond dimension $M=1$. Using this form, however, leads to large initial errors (stiffness), because the so-called left and right blocks~\cite{Schollwock-2011} do not contain the outgoing states generated by interaction terms. To resolve this issue, we used the same bond expansion technique already introduced in quantum chemistry to boost ground-state simulations~\cite{Legeza-2003, Barcza-2011}, where the blocks are extended by block states, orthogonal to the original, and are present in $\hat{H}^n \ket{\psi_0}$ for $n\in [ 1\dots n_{\max} ]$.

\subsection{Execution on IBM Quantum hardware} \label{subsec:methods_execution}

The Heron r2 processors consist of 156 fixed-frequency transmon qubits arranged in a heavy-hex lattice structure, each connected by tunable couplers. The results presented in this work were specifically obtained using \texttt{ibm$\_$kingston} and \texttt{ibm$\_$marrakesh}, which operate respectively at $250$K and $195$K circuit layer operations per second (CLOPS) \cite{wack2021} and show an averaged two-qubit error rate per layered two-qubit gate (EPLG) of $9.93\times 10^{-4}$ and $3.98\times 10^{-3}$\cite{mckay2023benchmarkingquantumprocessorperformance}. Fig.~\ref{fig:supp_dev} of the extended data details the characteristics of the device recorded at the time of execution of the results shown in the main text. Here, each vertex in the connectivity graph represents the T2 coherence time, and the connecting edges depict the error rate of the native two-qubit gate (CZ). Table~\ref{tab:calibration_data} shows the median decoherence times and gate error rates at the time of execution for the different quantum simulations. 

The quantum circuits for each simulation are executed using Qiskit Runtime’s \texttt{Batch} execution mode, which is specifically designed to handle experiments consisting of multiple independent jobs efficiently. This mode enables classical computations, including pulse-level compilation, to be performed in parallel, thereby reducing waiting times between circuit executions and reducing drift effects. To obtain measurement outcomes, we sample bitstrings with the \texttt{SamplerV2} Qiskit primitive, which enables the direct implementation of error suppression techniques such as Pauli twirling~\cite{javadiabhari2024quantumcomputingqiskit}. Access to raw bitstrings allows for the customized reconstruction of the expectation values of the observables of interest, making it possible to implement bespoke error mitigation strategies, such as Gauss sector correction (GSC), operator decoherence renormalization (ODR) or gauge dynamical decoupling (GDD) (see Sect.~\ref{subsec:methods_em}).

For each time step, we execute two distinct circuits. First, we run a calibration circuit that is used to implement the ODR. Then, we execute the Trotter circuit for each time instant. Conveniently, the calibration and simulation circuits are executed consecutively to minimize potential drifting effects, although we have found that drifting is not the most relevant issue for the duration of a single simulation. The execution time of the batches we run is $1\mathrm{h}\,10\mathrm{min} \, \pm 5\mathrm{min}$. These batches contain $68$ different circuits ($34$ Trotter time instants multiplied by a factor of two when accounting for calibration circuits), which are spread over 136 jobs to reduce overhead in Qiskit Runtime and increase stability. In IBM's quantum hardware, the time of execution for the \texttt{SamplerV2} Qiskit primitive is almost entirely determined by the number of shots that we use to estimate observables, since measurement and qubit reset are the most costly operations in terms of time of execution. In all simulations, we run $300000$ shots per circuit, that is, $600000$ shots per time instant. Note that not all the shots contribute to the final expectation value, as we discard some of them based on the number of bit-flips detected by the decoder as outlined in Sect.~\ref{subsubsec:methods_em_gse}. To compute expectation values, we always maintain a threshold of $30000$ shots.

Hardware-level optimization is performed using Qiskit's preset pass manager with optimization level $3$, which applies a sequence of transpilation and noise-aware transformations tailored to the target device.  An instance of the Qiskit's preset pass manager defines a full compilation pipeline, which includes the following steps: \textit{init}, \textit{layout} -  to map the virtual qubits in the circuit to the physical qubits on the backend, \textit{routing} - which inserts SWAP gates to match the topology of the device, \textit{translation} - to translate to the target backend’s native basis gate set, and finally \textit{optimization}, to reduce the cost of executing the circuit, and \textit{scheduling}, as a final hardware-aware pass that schedules all the operations. As this process is not deterministic, two similar circuits can result in different qubit layouts after compilation. To ensure consistency across a batch of circuits, we fix the layout used for the first circuit and apply it to all subsequent circuits in the batch. For large simulations that involve more than $~40$ qubits, the probability of including a faulty qubit in the simulation becomes significant. To mitigate the impact of outliers, an effective strategy is to average the results over multiple layouts. In that case, the layouts are manually selected, taking into account the specific noise characteristics of the device. This approach virtually distributes the influence of the faulty qubit across the lattice, improving the accuracy of local observables. However, when all qubits in the selected layout are functioning properly, averaging over multiple layouts may introduce unnecessary variability.

To suppress coherent noise during execution, we enable Pauli twirling, also known as randomized compiling~\cite{randomized_compiling}, which is an error suppression technique available in Qiskit Runtime. Pauli twirling works by inserting randomized Pauli gates before and after certain two-qubit entangling operations in such a way that the ideal logical operation is preserved, while coherent errors are converted into stochastic Pauli errors. This transformation makes the noise more amenable to the error mitigation and correction techniques introduced in Sect.~\ref{subsec:methods_em}. In practice, twirling is activated through Qiskit Runtime, and it incurs no additional quantum resources beyond the inserted single-qubit gates. In particular, measurement results are averaged over multiple randomizations within the $300000$ executions per circuit that we have fixed manually.

\begin{table}[]
\centering
\begin{tabularx}{0.9\textwidth}{ 
| >{\centering\arraybackslash}X |
| >{\centering\arraybackslash}X 
| >{\centering\arraybackslash}X 
| >{\centering\arraybackslash}X 
| >{\centering\arraybackslash}X |
| >{\centering\arraybackslash}X 
| >{\centering\arraybackslash}X |}
\hline
Simulation $(\mathrm{Size}, m, g)$ & Device & Readout error & Relaxation time $(T_1)$ & Dephasing time $(T_2)$ & $SX$ error & $CZ$ error \\ \hline \hline \vspace{1pt}
$(2\times2, 5, 2)$ 1-string & \texttt{ibm\_kingston} & $8.67 \cdot 10^{-3}$ & $248.74\,\mu\mathrm{s}$ & $146.95\,\mu\mathrm{s}$ & $2.31 \cdot 10^{-4}$ & $1.95\cdot10^{-3}$ \\ \hline \vspace{1pt}
$(2\times2, 5, 2)$ 3-string & \texttt{ibm\_marrakesh} & $1.31 \cdot 10^{-2}$ & $195.18 \,\mu\mathrm{s}$ & $89.01\,\mu\mathrm{s}$ & $2.77\cdot10^{-4}$ & $3.37\cdot10^{-3}$ \\ \hline \vspace{1pt}
$(3\times3, 5, 0.01)$ & \texttt{ibm\_kingston} & $7.67 \cdot 10^{-3}$ & $211.09\,\mu\mathrm{s}$ & $122.72\,\mu\mathrm{s}$ & $2.74 \cdot 10^{-4}$ & $1.99\cdot10^{-3}$ \\ \hline \vspace{1pt}
$(3\times3, 0.3, 0.5)$ & \texttt{ibm\_kingston} & $8.54 \cdot 10^{-3}$ & $276.32\,\mu\mathrm{s}$ & $156.33\,\mu\mathrm{s}$ & $2.31 \cdot 10^{-4}$ & $1.95\cdot10^{-3}$ \\ \hline \vspace{1pt}
$(3\times7, 0, 0)$ & \texttt{ibm\_marrakesh} & $1.66 \cdot 10^{-2}$ & $93.57\,\mu\mathrm{s}$ & $188.89\,\mu\mathrm{s}$ & $3.14 \cdot 10^{-4}$ & $2.57\cdot10^{-3}$ \\ \hline
\end{tabularx}
\caption{\justifying Median IBM's quantum hardware calibration data at the moment of execution of the different simulations.}
\label{tab:calibration_data}
\end{table}

\subsection{Bootstraping for error bars} \label{boots}

Bootstrapping is a nonparametric statistical technique that is used to estimate the uncertainty of a parameter, such as the mean, median, or fit coefficient, by resampling data with replacement. It is particularly powerful when the underlying distribution is unknown or when the propagation of analytic errors is intractable.

The method begins with a dataset of size $N$. From this, a large number $B$ of "bootstrap samples" are generated by randomly sampling $N$ data points with replacement, meaning that individual data points may appear more than once in a single sample. For each bootstrap sample, the statistic of interest is computed, resulting in a distribution of $B$ bootstrap estimates. This empirical distribution approximates the sampling distribution of the statistic.

The error bars are then derived by computing the standard deviation of the bootstrap distribution (yielding an estimate of the standard error) or by constructing confidence intervals, for instance, using the percentile method, where the lower and upper bounds correspond to specific quantiles (e.g., the 2.5 and 97.5 percentiles for a 95\% confidence interval).

Bootstrapping makes minimal assumptions about the form of the data distribution, relying only on the assumption that the sample is representative of the population. It is computationally intensive, but widely applicable, especially in modern scientific contexts where analytic solutions are either biased or unavailable.

\newpage

\subsection{Extended data}

\begin{figure*}[!ht]
\centering
\includegraphics[width=0.99\textwidth]{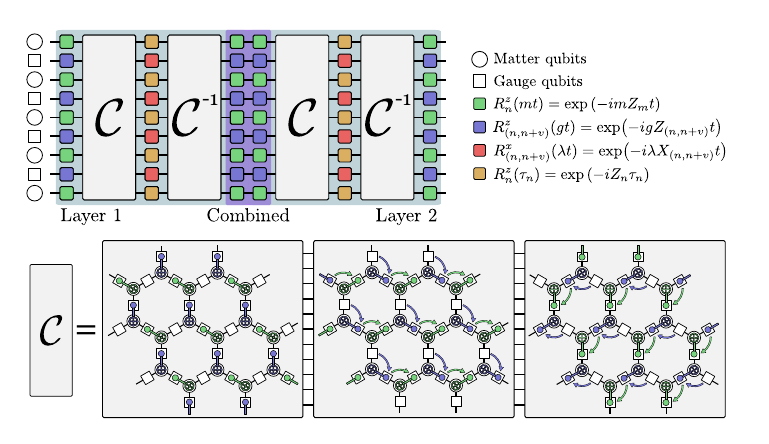}
\caption{\justifying Trotter circuits for the quantum simulation of the $Z_2$HM. This circuit is designed by repeated composition of the Pauli gadget drawn in Fig.~\ref{fig:fig0} in a specific order (Sect.~\ref{subsec:methods_trotter}) so that the CNOTs commute and they can be combined in blocks $\mathcal{C}$ of depth 3. We divide the matter sites on the lattice into two sets, coloured in blue and green. We place the Pauli gadgets acting on the qubits in each set in parallel, and the sets act sequentially.}
\label{fig:trotter_circuits}
\end{figure*}

\begin{figure*}[!ht]
\includegraphics[width=0.99\textwidth]{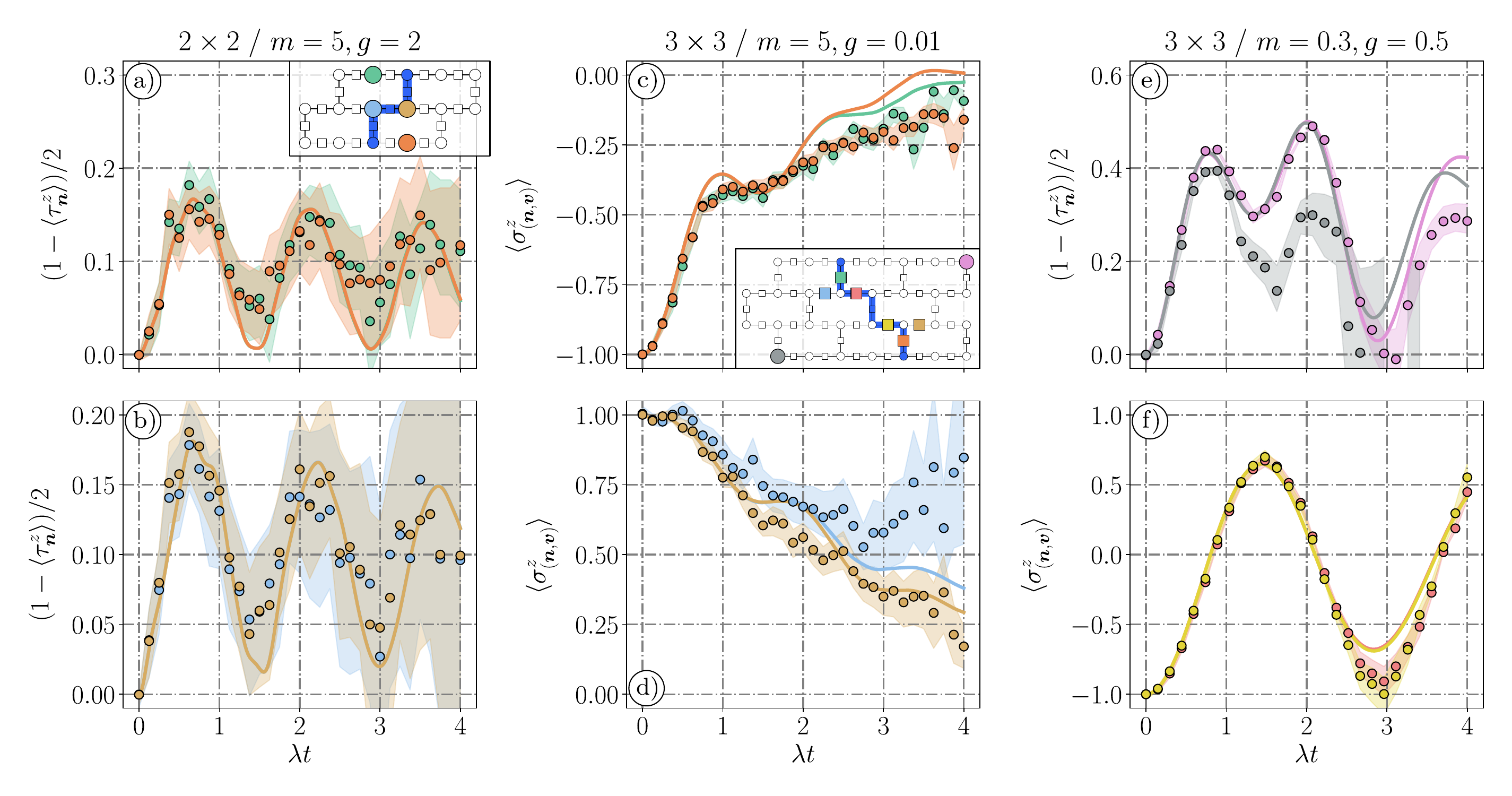}
\caption{\justifying Additional data for the experiments in Fig.~\ref{fig:single_string_dynamics}. (a)-(b) show the transfer of population from the initial endpoints of the string to neighbouring sites in the confined phase. We exclusively observe yo-yo oscillations, also in the matter sites interior to the string, as the bending mode of motion only virtually populates the states with a shorter string. (c)-(d) features the local expectation value of the gauge field at the colored links, interior and exterior to the initial string, respectively. This expectation value approaches a steady value around $\langle \sigma_{(\boldsymbol{n}, \boldsymbol{v})}^z \rangle \simeq 0$, indicating that the gauge field tends to a superposition of every possible configuration in the 2-particle sector. In (e), we plot the dynamics of the occupation in two bivalent nodes of the lattice in the Higgs phase. The grey points do not adjust to the MPS curve because the qubit mapped to the grey site was faulty at the time of the simulation. The propagation of errors introduced by this qubit is well controlled enough that its effect is negligible on the other sites considered. (f) shows the local expectation value of the gauge field in the Higgs phase.}
\label{fig:supp_1string_otherp}
\end{figure*}

\begin{figure*}[!ht]
\includegraphics[width=0.99\textwidth]{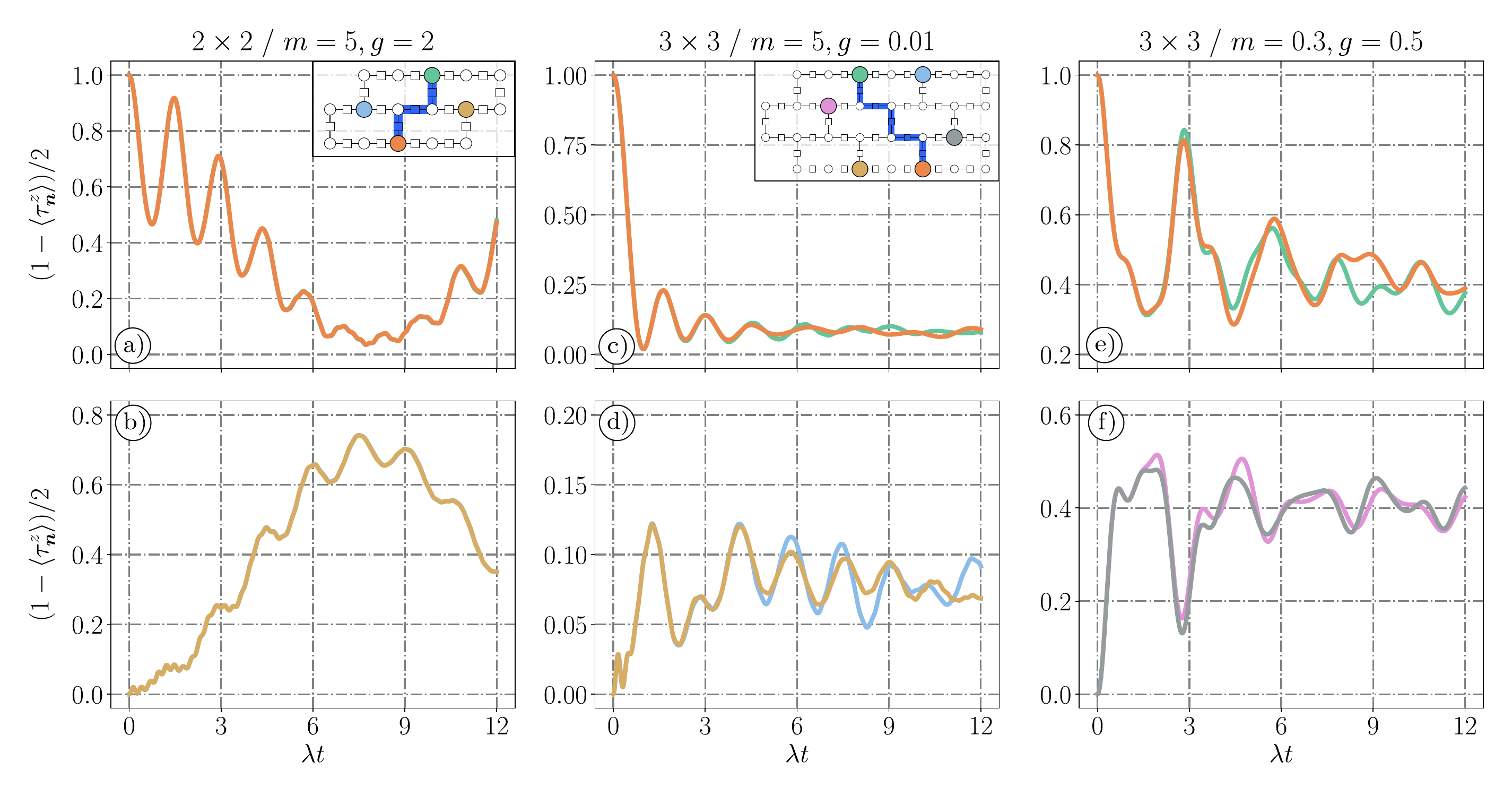}
\caption{\justifying Results from MPS-based simulations of the single string quench for long times. In (a-b), we plot the occupation at the initial and rotated string endpoints in the confined phase. Green and blue curves overlap with the orange and yellow ones. The longer times accessible to the MPS simulations enable the resolution of more than half oscillation bending mode, corresponding to the high-amplitude oscillation that accompanies the high-frequency yo-yo motion. The occupation in the initial endpoints (a) almost completely vanishes for $t=T_\mathrm{b}/2 \simeq 7.54\lambda^{-1}$, signaling that the string has rotated almost completely. The yo-yo motion is also manifest in the rotated string as the small-amplitude oscillations around $t=T_\mathrm{b}/2$ in (b). (c-d) feature the occupation of initially occupied and empty matter sites, respectively. In the deconfined phase, the system quickly reaches a long-lived steady state with the initially present matter particles spread on the lattice and the gauge field in a superposition of all the different configurations allowed by the gauge symmetry. Revivals are expected at a timescale proportional to the size of the lattice $t_\mathrm{R} \sim \mathcal{O}(N)$. (e-f) show the occupations in the Higgs phase, where long-lived glassy damped oscillations are observed. For small $m$, these oscillations couple the initial state $\ket{\psi_0}$ with $H_I\ket{\psi_0}$, which is only possible if creating matter from the vacuum is energetically cheap. As $m$ increases, spreading is favored over vacuum oscillations, and the glassy oscillations dampen. These MPS-based simulations have been computed using the ``basis update and Galekin'' integrator.}
\label{fig:supp_1string_tn}
\end{figure*}

\begin{figure*}[!ht]
\includegraphics[width=0.6\textwidth]{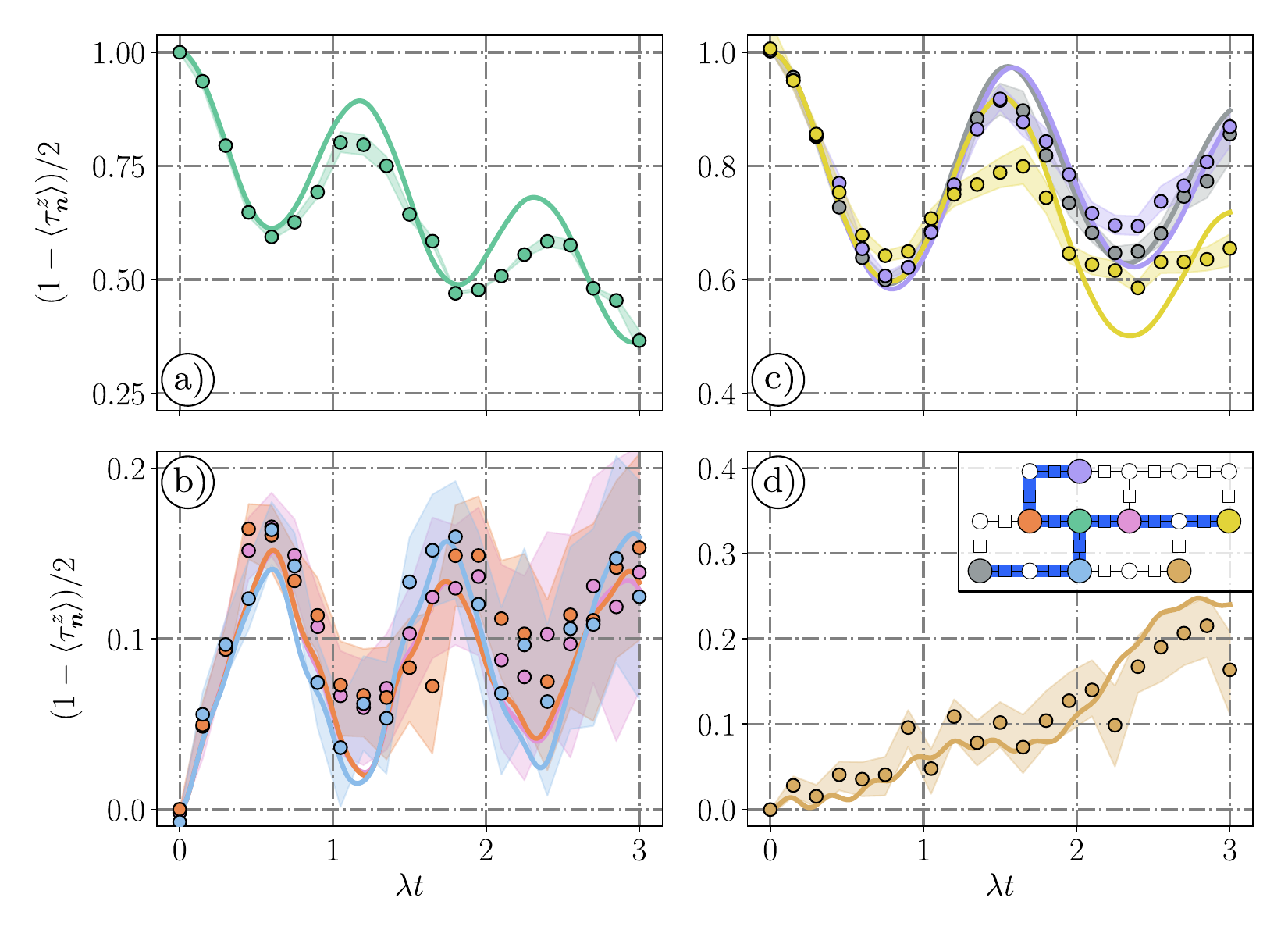}
\caption{\justifying Single site occupations for the 3-string dynamics. The particle initially placed in the green site, the occupation of which appears in (a), first tunnels out to the neighbouring sites displayed in (b) in a first-order process similar to the yo-yo motion. However, second-order processes prevent it from completely tunneling back to its initial position, and the population in (a) experiences a steady decline that does not translate to an increase in the neighbouring sites of (b). These effects combined translate into the population peak of the broken string configurations plotted in Fig.~\ref{fig:double_string_dynamics} of the main text. In (c), we plot the occupation at the initial string endpoints on the boundary of the lattice. The motion of the particles initially in the purple and gray sites is restricted to neighbouring positions for the times considered in the simulation. The particle initially in the light-yellow site experiences a yoyo motion, and that string endpoint rotates into the dark-yellow site.}
\label{fig:supp_3string}
\end{figure*}

\begin{figure*}[!ht]
\includegraphics[width=\textwidth]{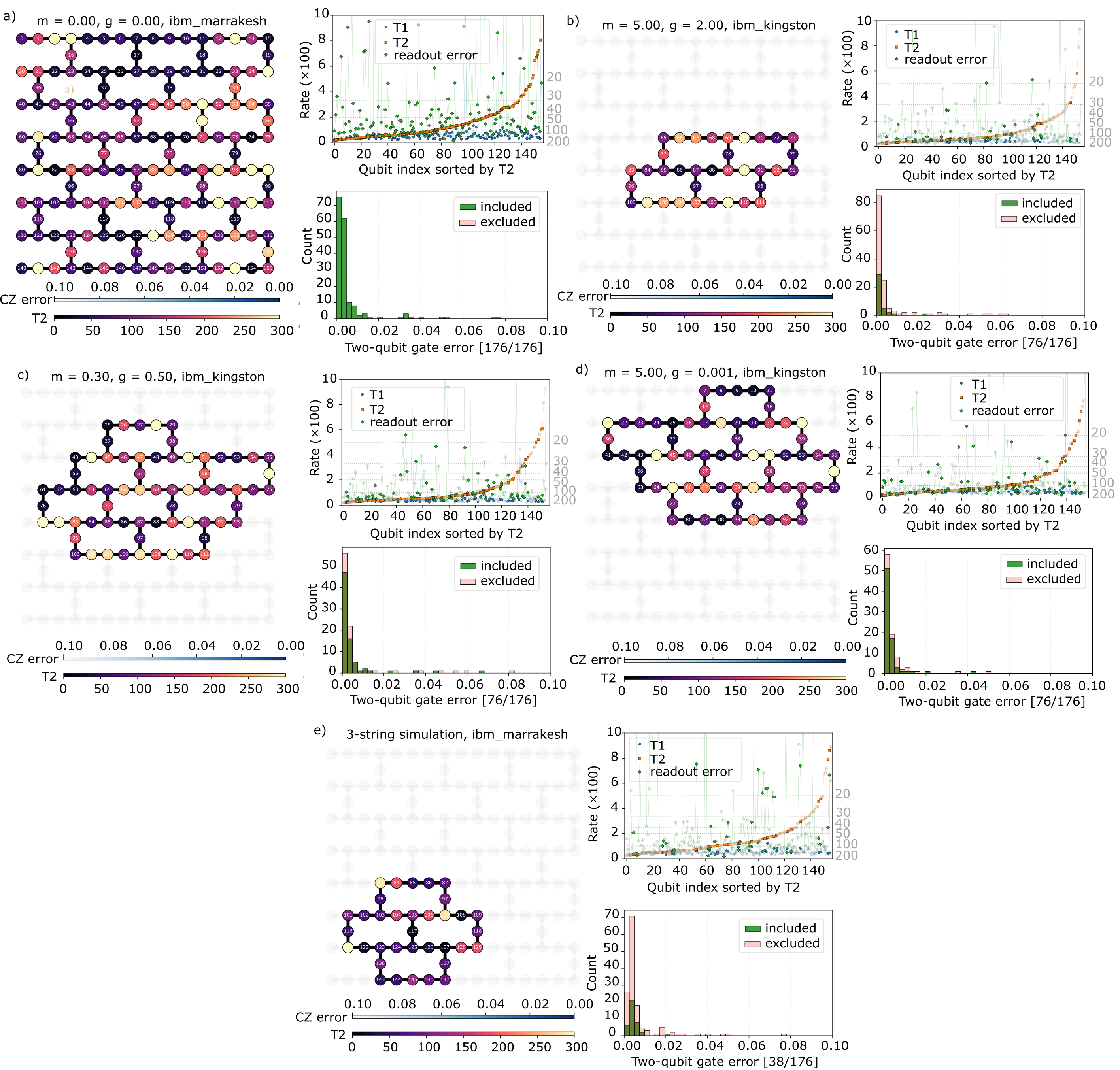}
\caption{\justifying Device characterization for the experiments presented in Fig.\ref{fig:error_mitigation_data}(a) in panel (a), Figs.~\ref{fig:single_string_dynamics}(a)-(b) in panel (b),  Figs.~\ref{fig:single_string_dynamics}(e)-(f) in panel (c), Figs.~\ref{fig:single_string_dynamics}(c)-(d) in panel (d) and Figs.~\ref{fig:double_string_dynamics}(a)-(b) in panel (e). For each experiment, we display the qubit-wise CZ gate error on the device connectivity graph and the individual qubit T2 dephasing times, with the specific qubit layout used in each experiment highlighted. Note that the device layouts are isomorphic representations of the simulation layouts illustrated in the corresponding figures. Additionally, we show the T1 relaxation times and readout errors per qubit, as well as a histogram of two-qubit (CZ) gate errors for the qubits included in each experiment.}
\label{fig:supp_dev}
\end{figure*}
\end{document}